\UseRawInputEncoding 
\documentclass[pra,aps,amssymb,twocolumn,noshowpacs]{revtex4-1}
\usepackage{color}
\usepackage{graphicx}
\usepackage{hyperref}
\usepackage{amsmath}
\usepackage{latexsym}
\usepackage{amssymb}
\usepackage{mathrsfs}
\usepackage{mathtools}
\usepackage{textcomp}
\usepackage{layout}
\usepackage{verbatim}
\usepackage{bm}
\usepackage{amsfonts,epsfig}

\begin{document}

\title{Single-site Rydberg addressing in 3D atomic arrays for quantum computing with neutral atoms}

\date{\today}
\author{Xiao-Feng Shi}
\affiliation{School of Physics and Optoelectronic Engineering, Xidian University, Xi'an 710071, China}

\begin{abstract}
  Neutral atom arrays are promising for large-scale quantum computing especially because it is possible to prepare large-scale qubit arrays. An unsolved issue is how to selectively excite one qubit deep in a 3D atomic array to Rydberg states. In this work, we show two methods for this purpose. The first method relies on a well-known result: in a dipole transition between two quantum states driven by two off-resonant fields of equal strength but opposite detunings $\pm\Delta$, the transition is characterized by two counter-rotating Rabi frequencies $\Omega e^{\pm i\Delta t}$~[or $\pm\Omega e^{\pm i\Delta t}$ if the two fields have a $\pi$-phase difference]. This pair of detuned fields lead to a time-dependent Rabi frequency $2\Omega \cos(\Delta t)$~[or $2i\Omega \sin(\Delta t)$], so that a full transition between the two levels is recovered. We show that when the two detuned fields are sent in different directions, one atom in a 3D optical lattice can be selectively addressed for Rydberg excitation, and when its state is restored, the state of any nontarget atoms irradiated in the light path is also restored. Moreover, we find that the Rydberg excitation by this method can significantly suppress the fundamental blockade error of a Rydberg gate, paving the way for a high-fidelity entangling gate with commonly used quasi-rectangular pulse that is easily obtained by pulse pickers. Along the way, we find a second method for single-site Rydberg addressing in 3D, where a selected target atom can be excited to Rydberg state while preserving the state of any nontarget atom due to a spin echo sequence. The capability to selectively address a target atom in 3D atomic arrays for Rydberg excitation makes it possible to design large-scale neutral-atom information processor based on Rydberg blockade.

\end{abstract}

\maketitle

\section{introduction}
Neutral atom arrays have become a promising platform for large-scale quantum computing~\cite{PhysRevLett.85.2208,Saffman2010,Saffman2011,Saffman2016,Weiss2017,Adams2019}. The attraction lies in that one can not only prepare large-scale qubit arrays and initialize the state for the qubits with high fidelity, but can also prepare high-fidelity single-qubit gate and easily read out the quantum information stored in the qubits~\cite{PhysRevLett.103.153601,RSI2014,Nogrette2014,Xia2015,Zeiher2015,Ebert2015,Wang2016,Barredo2016}. The fidelity of two-qubit entangling gates is growing steadily recently~\cite{Wilk2010,Isenhower2010,Zhang2010,Maller2015,Zeng2017,Picken2018,Levine2018,Graham2019,Levine2019}, from the initial value of $73\%$~\cite{Isenhower2010} to $97\%$~\cite{Levine2019}, which points to the likeliness that high-fidelity universal gate set can be experimentally realized in the near future for the required accuracy of measurement-free fault-tolerant quantum computing~\cite{Crow2016}. Nevertheless, there are still many unsolved issues toward this goal as reviewed in Ref.~\cite{Saffman2016}. 

One issue for large-scale quantum information processing in a 3-dimensional~(3D) qubit array is how to selectively excite one qubit deep in the lattice between ground and Rydberg states. To date, there has been demonstration of two-qubit gate in a two-dimensional~(2D) array of neutral atoms~\cite{Maller2015,Graham2019}, where the single-site addressing without affecting nontarget atoms is guaranteed by sending laser fields perpendicular to the 2D array, and thus is not applicable in a 3D lattice. There have been several experiments on single-qubit gates in 3D~\cite{Wang2015,Wang2016} or 2D~\cite{Xia2015} atomic arrays by using a hybrid irradiation of laser and microwave fields. These methods depend on shifting the atomic transition frequency by sending laser fields upon target atoms, where the Stark shifts~(divided by the Planck constant) should be comparable to the inverse of the gate durations of several hundred $\mu s$. To the best of our knowledge, however, there is no method of single-site Rydberg addressing in a 3D lattice where nontarget atoms in the light path are not influenced.

In this work, we present two methods for selectively exciting one qubit in any site of a 3D qubit array to Rydberg state. The first method relies on a well-known phenomenon: the action of two symmetrically detuned laser pulses is equivalent to that of a monochromatic field whose amplitude is sinusoidally modulated in time, as studied in Ref.~\cite{Goreslavsky1980}. This means that a full dipole transition can also occur by absorbing two fractional photons, i.e., by absorbing half of a photon with energy $E_{ge}+\hbar\Delta$, and half of another photon with energy $E_{ge}-\hbar\Delta$, where $\hbar$ and $\Delta$ are the reduced Planck constant and frequency detuning, respectively. This resonance can be called off-resonance-induced resonance~(ORIR). By ORIR, one laser light is sent along one direction, while the other sent along another direction, and they together excite a target atom. The nontarget atom can accumulate phase shift or even experience state evolution in general, but we show that our ORIR-based theory can eliminate its effect when the state of the target atom is restored. Besides the ORIR-based method, we show a second method with a three-level ladder type system, where one target atom can be excited to Rydberg state while preserving the state of any nontarget atom. In the second method, the state of the target atom can pick a $\pi$ phase shift upon the completion of the ground-Rydberg-ground state transfer; such a $\pi$ phase is crucial in the controlled-phase gate based on Rydberg blockade. These methods make it possible to couple only one atom to the Rydberg state in 3D.

We further show that ORIR can lead to high fidelity in the Rydberg blockade gate. In quantum computing with neutral atoms and Rydberg interactions~\cite{PhysRevLett.85.2208,Saffman2010,Saffman2011,Saffman2016,Weiss2017,Adams2019}, it has been an outstanding challenge to design a practical high-fidelity entangling gate~\cite{Goerz2014,Theis2016,Shi2017,Petrosyan2017,Shi2018prapp2,Shen2019,Shi2019,Yu2019,Levine2019}. A traditional method to achieve an entangling gate by Rydberg interactions is via the blockade mechanism~\cite{PhysRevLett.85.2208}, in which there is a fundamental blockade error. We show that when the Rabi frequency $\Omega$ in the usual method is replaced by an ORIR-induced Rabi frequency $i\Omega\sin(\Delta t)$, the blockade error can be suppressed by more than two orders of magnitude. More important, this reduction of error by ORIR is robust against the variation of the blockade interaction, which compares favorably to other pulse-shaping-based methods for suppressing the blockade error. Thus, ORIR can effectively remove the blockade error, making it possible to realize a high-fidelity neutral-atom entangling gate with quasi-rectangular pulses that are easily attainable by pulse pickers.

The remainder of this work is organized as follows. In Sec.~\ref{sec02}, we give details about how ORIR appears. In Sec.~\ref{sec03}, we study two methods for single-site Rydberg addressing in a 3D optical lattice; a comparison between the two methods is given at the end of Sec.~\ref{sec03}. In Sec.~\ref{sec04b}, we show that ORIR can suppress a fundamental rotation error in the Rydberg blockade gate. In Sec.~\ref{sec05}, we discuss other applications of ORIR. Section~\ref{sec06} gives a brief summary.


\section{Resonance from off resonance}\label{sec02}

\begin{figure}
\includegraphics[width=3.4in]
{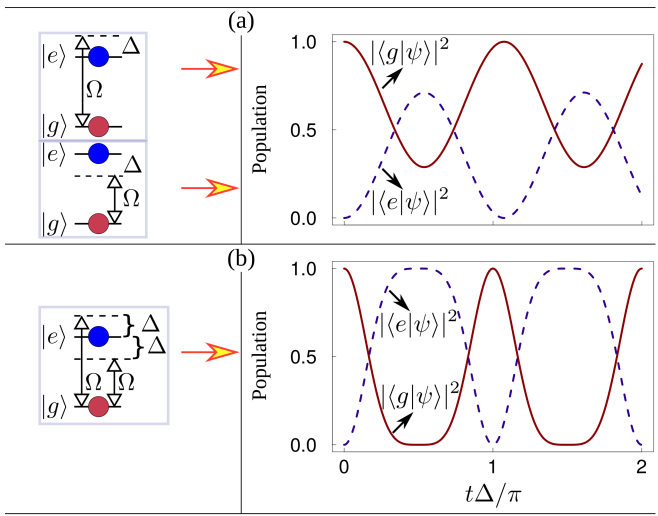}
 \caption{(a) Schematic and population dynamics of Rabi oscillations between $|g\rangle$ and $|e\rangle$ driven by an external field with a detuning of $\Delta$ or $-\Delta$. The initial state is $|\psi(0)\rangle=|g\rangle$. (b) When the transition $|g\rangle\leftrightarrow|e\rangle$ is driven by two external coherent fields with opposite detunings $\pm\Delta$, an effective Rabi frequency of $2\Omega\cos(\Delta t)$ appears. This leads to a resonant transition between $|g\rangle$ and $|e\rangle$. The condition  $\Delta/\Omega=2/\pi$ is used in both (a) and (b). \label{figure01} }
\end{figure}

The reason for ORIR to occur lies in that the dipole coupling between an atom and two symmetrically detuned laser fields can be characterized by a sinusoidal Rabi frequency $2i\Omega \sin(\Delta t)$ or $2\Omega \cos(\Delta t)$~\cite{Goreslavsky1980}. To put it in perspective we consider a pair of external coherent electromagnetic fields of equal strength but with opposite detunings $\pm\Delta$ applied for the dipole transition $|g\rangle\leftrightarrow|e\rangle$. In the dipole approximation, these two fields lead to a pair of counter-rotating Rabi frequencies $\Omega_\pm=\Omega e^{\pm i\Delta t}$. When only one of these two detuned fields is applied, the expectation value of the population in $|e\rangle$ has an upper bound $\Omega^2/(\Omega^2+\Delta^2)$ if the initial state is $|g\rangle$~\cite{Shi2017,Shi2018prapp2}, as shown in Fig.~\ref{figure01}(a). But when the two oppositely detuned fields are applied simultaneously and because $\Omega_++\Omega_-=2\Omega \cos(\Delta t)$, a full transition between $|g\rangle$ and $|e\rangle$ becomes attainable within a time of less than $|\pi/\Delta|$ as long as $|\Omega/\Delta|\geq \pi/2$. The population dynamics in the two-level system in response to the detuned driving is shown in Fig.~\ref{figure01} when $\Omega/\Delta= \pi/2$. Figures~\ref{figure01}(a) and~\ref{figure01}(b) show that the dynamics in the two-level system driven by the two oppositely detuned fields dramatically differs from that when only one detuned driving is present. This phenomenon results from the quantum interference between the two Rabi oscillations with opposite detunings. Remarkably, the speed of this off-resonance-induced resonant transition is comparable to its resonant counterpart: when $|\Omega/\Delta|=\pi/2$, a full transition between the two states is achieved within a time of $t_{\pi}\pi/2$ that is comparable to the duration, $t_{\pi}=\pi/(2\Omega)$, of a $\pi$ pulse required for a full transition in the resonant case with the Rabi frequency $2\Omega$. This type of multi-photon ORIR can occur in various electric or magnetic dipole transitions of an atom or molecule, either natural or artificial.

We give a detailed mathematical argument for the above discussion. For a two-level system in Fig.~\ref{figure01}, the eigenenergies of $|g\rangle$ and $|e\rangle$ are zero and $E_{\text{ge}}$, respectively. The electric dipole transition between $|g\rangle$ and $|e\rangle$ is driven by two laser fields, one with central (angular) frequency $\omega+\Delta$, and the other with central frequency $\omega-\Delta$, where $\omega=E_{\text{ge}}/\hbar$. The Hamiltonian~(divided by $\hbar$) of the matter-light coupling in dipole approximation is
\begin{eqnarray}
  \hat{H}(t) &=&\omega |e\rangle\langle e|+[ \Omega_{1}  (e^{it(\omega+\Delta )} + e^{-it(\omega+\Delta )} ) |e\rangle\langle g|\nonumber\\
 &&+ \Omega_{2}(e^{it(\omega-\Delta )} + e^{-it(\omega-\Delta )} ) |e\rangle\langle g|+\text{H.c.} ]/2,\label{sec02equation01}
\end{eqnarray}
where we assume that $\Omega_1$ and $\Omega_2$ are real, and ``H.c.'' denotes the Hermitian conjugate. In principle, there is a phase difference $\varphi_0=(\mathbf{k}_1-\mathbf{k}_2)\cdot\mathbf{r}$ between $\Omega_1$ and $\Omega_2$, where the subscript $1(2)$ distinguishes the two transitions with opposite detunings. If the Rydberg state $|e\rangle$ is excited by couterpropagating fields along $\mathbf{z}$ via two-photon excitation through an intermediate state with a GHz-scale detuning $\delta_{1(2)}$, and if meanwhile the fields for $\mathbf{k}_1$ and $\mathbf{k}_2$ copropagate, $\varphi_0$ becomes $2(\Delta+\delta_2-\delta_1)z/c$, where $z$ is the z-component coordinate of the atom, $c$ the speed of light. The difference between $\delta_{1}$ and $\delta_{2}$ is much larger than $\Omega_{1(2)}$ so that a common intermediate state is used for the two two-photon transitions; for instance, $\delta_{1}$ and $\delta_{2}$ can have different signs. Because $\Delta\sim\Omega_{1(2)}$ and the effective two-photon Rabi frequency $\Omega_{1(2)}$ is on the order of MHz, $\varphi_0\approx2(\delta_2-\delta_1)z/c$. For atoms cooled to temperatures around $1$~K or below, the change of $z$ within several $\mu$s is on the order of $\mu$m, which means that $\varphi_0$ is almost constant; furthermore, $|\delta_2-\delta_1|/c\ll k_1,k_2$, and thus we can assume that this phase has been compensated by adjustment of the overall phases carried by the laser beams. This is technically possible since the phase fluctuation of a laser field can be made negligible~\cite{Saffman2005,Wineland1998}. But for single-site Rydberg addressing with our methods~(shown in Sec.~\ref{sec03}), the fields for $\Omega_1$ and those for $\Omega_2$ propagate along different directions, e.g., one along $\mathbf{z}$ and the other along $\mathbf{x}$, then $\varphi_0\approx \mathcal{K}(z-x)$, where $\mathcal{K}$ is the difference of the wavevectors of the upper and lower transitions. For configurations with $ \mathcal{K}$ of several $10^{6}m^{-1}$~\cite{Isenhower2010}, the fluctuation of the qubit locations lead to large errors in $\varphi_0$, and it is necessary to cool atoms to very low temperatures to establish ORIR.

Using the operator $\hat{\mathcal{R}}= \omega |e\rangle\langle e|$ for a rotating-frame transform, $e^{i\hat{\mathcal{R}}t }\hat{H}e^{-i\hat{\mathcal{R}}t } - \hat{\mathcal{R}}$, the Hamiltonian becomes,
\begin{eqnarray}
  \hat{H}(t) &=& \Omega_1 (e^{it(2\omega+\Delta )} + e^{-it\Delta } ) |e\rangle\langle g|/2\nonumber\\
 &&+ \Omega_2 (e^{it(2\omega-\Delta )} + e^{it\Delta } ) |e\rangle\langle g|/2+\text{H.c.}.\nonumber
\end{eqnarray}
When $\Omega_1=\Omega_2=\Omega$, the above equation simplifies to,
\begin{eqnarray}
  \hat{H}(t)  &=&  \Omega (e^{2it\omega} +1 )\cos(t\Delta) |e\rangle\langle g| + \text{H.c.}.\nonumber
\end{eqnarray}
We further assume $\Delta\sim|\Omega|\ll \omega$ so that the Bloch-Siegert shift is negligible~\cite{CohenTannoudji}. In this case, $e^{\pm2it\omega}$ is rapidly oscillating and can be discarded according to the rotating wave approximation. This leads to the following Hamiltonian,
\begin{eqnarray}
  \hat{H} (t) &=&  \Omega \cos(t\Delta) (|e\rangle\langle g| +|g\rangle\langle e|) .\nonumber
\end{eqnarray}
Starting from the initial state, $|\psi(0)\rangle =|g\rangle$, the system wavefunction $|\psi(t)\rangle = C_g(t) |g\rangle + C_e(t)|e\rangle$ evolves according to
\begin{eqnarray}
  C_g(t)&=&\cos \left[\frac{\Omega}{\Delta} \sin(t\Delta) \right],\nonumber\\
  C_e(t)&=&-i \sin \left[\frac{\Omega}{\Delta} \sin(t\Delta) \right].\label{CgCe}
\end{eqnarray} 
When $ \frac{|\Omega|}{\Delta} \geq \frac{\pi}{2}$ is satisfied, the transition probability from the ground state to the excited state can reach 1. A critical condition for a full transition is
\begin{eqnarray}
 \frac{|\Omega|}{\Delta} = \frac{\pi}{2},~~t=\frac{\pi}{2\Delta}.\nonumber
\end{eqnarray} 
Remarkably, the time for the transition from the ground state to the excited state is $\frac{\pi}{2\Delta}$, which is only $\pi/2$ times longer than the time $\frac{\pi}{2|\Omega|}$ of a $\pi$ pulse in a resonant transition with a Rabi frequency $2\Omega$.

ORIR is a useful method in quantum control with atomic ions~\cite{Sorensen1999}. Below, we show its applicability in neutral atoms.


\section{Single-site Rydberg addressing in 3D arrays}\label{sec03}
ORIR can be used for Rydberg addressing of a single atom in a 3D optical lattice. In general, the difficulty of Rydberg addressing in a 3D lattice lies in two aspects. First, sending lasers to a target atom in a dense 3D lattice can influence other atoms along the light path, and it is unlikely to leave the state of a nontarget atoms intact after fully Rydberg exciting and deexciting a target atom. Second, the atoms along the light path can exhibit Rydberg blockade, bringing the problem a many-body complexity. One may image that for a two-photon excitation of $s$- or $d$-orbital Rydberg atoms, the two laser beams can be sent along different directions; because both laser fields are largely detuned, only the target atom at the intersection of the two beams are Rydberg excited. However, phase twist can occur for the nontarget atom even if it is pumped only by an off-resonant field~[see, for example, the dashed curve in Fig.~3(b) of Ref.~\cite{Shi2019}]. So, there can be a phase shift to the atomic state of a nontarget qubit illuminated by the laser fields addressing the lower transition in our problem. In fact, even if the detuning of the laser field addressing the lower transition is 10 times larger than its Rabi frequency $\Omega_{\text{g}}$, the phase shift to the ground state can reach $\pi/2$ if the atom is irradiated by a time of $20\pi/\Omega_{\text{g}}$. These issues can be tackled by using ORIR-based optical spin echo in a very small lattice, and a microwave spin echo assisted by ORIR in a relatively large one. Here, the optical spin echo is for restoring the state of nontarget atoms if no Rydberg interaction exists between the nontarget atoms, while the microwave spin echo is for removing many-body effect if there is Rydberg blockade in the nontarget atoms.

\begin{figure}
\includegraphics[width=3.2in]
{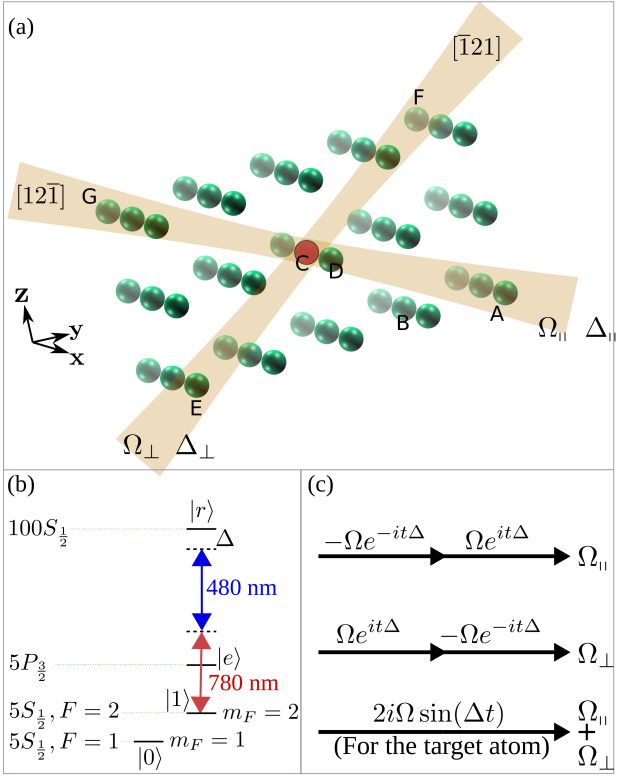}
\caption{(a) Diagram of targeting one atom by two detuned laser beams in a 3D lattice of rubidium atoms. Origin of the coordinate system is at the center of the target site. The quantization axis is along $[101]$; the two light paths are along $\mathbf{l}_\perp= [\overline{1}21]$ and $\mathbf{l}_\shortparallel=[12\overline{1}]$, respectively. The red atom labeled as ``C'' is the target atom. Atom $E$ and atom $F$~(atom $A$ and atom $G$) are nontarget atoms at the beam axis of fields along $\mathbf{l}_\perp$~($\mathbf{l}_\shortparallel$). (b) Atomic transition diagram. Both the light fields propagating along $\mathbf{l}_\perp$ and $\mathbf{l}_\shortparallel$ are polarized along the quantization axis, inducing the same transition. The detuning for the transition $|1\rangle\rightarrow|e\rangle$ is much larger than the two-photon detuning $\Delta$. (c) Time-dependence of the Rabi frequencies for method I. The effective Rabi frequency is $2i\Omega\sin(\Delta t)$ at the target atom.    \label{figure10} }
\end{figure}

We take a system shown in Fig.~\ref{figure10} for illustration. For a small array, the laser spot is small enough so that only an atom at the beam axis can be irradiated, and the laser intensity is of similar magnitude at the target atom and any nontarget atom along the light path. So, the Rabi frequencies are of similar magnitude for all the irradiated atoms in the problem. In Fig.~\ref{figure10}, a cubic lattice with $3\times5\times3$ sites is shown, the lattice constant is $\mathcal{L}$, and the relevant transition is between a hyperfine state of $F=2$ and a high-lying $s$-orbital Rydberg state of rubidium-87. The quantization axis is along $[101]$ which can be specified by a magnetic field. All laser fields are $\pi$-polarized along the quantization axis. We want to excite the central atom~(denoted by the red ball), which is located at the origin of the Cartesian coordinate~(the $\mathbf{x}-\mathbf{y}-\mathbf{z}$ arrows are only for clarifying the directions). In a rotating-frame as in Sec.~\ref{sec02}, a two-photon transition $|1\rangle\leftrightarrow|r\rangle$ with Rabi frequency $\Omega_\perp =\Omega e^{i(t\Delta_\perp+\varphi_\perp)}$ is created by sending focused Gaussian beams along $\mathbf{l}_\perp= [\overline{1}21]$, with the foci at the center of the target site. Here, $\varphi_\perp $ is determined by the phases of the laser oscillators and the distance that the light travels from the laser sources to the target atom; for brevity, we write $\varphi_\perp=\mathbf{k}_\perp\cdot\mathbf{r}$, where $\mathbf{k}_\perp$ is the wavevector of the fields. Furthermore, $\Delta_\perp$ is the overall detuning of the two-photon transition between the ground state $|1\rangle$ and the Rydberg state $|r\rangle$, which should be much smaller than the detuning $\delta$ for the transition $|1\rangle\rightarrow|e\rangle$. In our method, $\Delta$ is several MHz, thus $\delta$ should be at least several hundred MHz. Meanwhile, another set of lasers are sent along $\mathbf{l}_\shortparallel=[12\overline{1}]$ which also focus at the same target atom. In method I, the lasers along $\mathbf{l}_\perp$ and $\mathbf{l}_\shortparallel$ are of similar wavelengths, but with some difference so that the two two-photon transitions are built via very different one-photon detunings at the intermediate state. In the same rotating frame, the lasers along $\mathbf{l}_\shortparallel$ drive the transition $|1\rangle\leftrightarrow|r\rangle$ by a Rabi frequency $\Omega_\shortparallel =\Omega e^{i(t\Delta_\shortparallel+\varphi_\shortparallel)}$.

Our scheme requires the condition of $e^{i\varphi_\perp},~e^{i\varphi_\shortparallel}=\pm1$. Because both $\varphi_\perp$ and $\varphi_\shortparallel$ are determined by the laser sources and the length of the light path, it is possible to tune the laser phases to the condition of $e^{i\varphi_\perp},~e^{i\varphi_\shortparallel}=\pm1$ if the position fluctuation of qubits is negligible. For optically cooled qubits before each experimental cycle, there is fluctuation of the qubit positions, leading to fluctuation in $\varphi_\perp$ and $\varphi_\shortparallel $. For the configuration in Fig.~\ref{figure10}(b), we have $|\mathbf{k}_\perp|=|\mathbf{k}_\shortparallel|\approx 5\times10^6m^{-1}$ when the upper and lower fields propagate oppositely. This means that the fluctuation of the qubit position along the light path should be much smaller than $1~\mu$m to validate the method. In the experiment of Ref.~\cite{Graham2019}, the transverse~(longitudinal) position fluctuation of the qubit is $0.27~(1.47)~\mu$m, which will lead to large fluctuation of $\varphi_\perp$ and $\varphi_\shortparallel$ if similar traps are used here. Thus it is necessary to use sufficiently deep traps and efficient cooling. This is why the methods shown below can not be implemented by sending one laser field along $\mathbf{l}_\perp$ for a one-photon transition $|1\rangle\rightarrow|e\rangle$, another field along $\mathbf{l}_\shortparallel$ for the other one-photon transition $|e\rangle\rightarrow|r\rangle$, where $|e\rangle$ is the intermediate state shown in Fig.~\ref{figure10}(b). This is because the wavevector for these two one-photon transitions are much larger, which makes the condition of $e^{i\varphi_\perp}=e^{i\varphi_\shortparallel}=\pm1$ even more challenging to be satisfied. 

An alternative solution to the above issue is to use naturally existing transitions with negligible wavevector. For instance, the Rydberg excitation via two counterpropagating fields for $6^{1}S_1\rightarrow 6^{1}P_1\rightarrow   n^{1}S_0$~(or $n^{1}D_2$) of ytterbium suffers from a negligible Doppler dephasing because $k\lesssim10^{5}m^{-1}$~(see, for example, Fig.1 of Ref.~\cite{Lehec2018}). In this latter case, it is necessary to store quantum information in the nuclear spin states of, e.g., $^{171}$Yb~(or $^{173}$Yb) for the purpose of quantum computing with Rydberg interaction. For Rydberg excitation of one of the two nuclear spin qubit states, one can first put the other nuclear spin qubit state to the metastable excited state $^{3}P_0$ to avoid leakage. However, it is beyond the scope of this work to give detail about this latter scheme of Rydberg gate based on $^{171}$Yb. We assume that the fluctuation of $\varphi_\perp$ and $\varphi_\shortparallel$ has been suppressed.

Below, Secs.~\ref{sec03A} and~\ref{sec03B} show two optical spin-echo methods for the excitation of target atoms, termed as method I and method II. The above discussion is applicable to method I. In method II, the lasers along $\mathbf{l}_\perp$ is for the transition $|1\rangle\leftrightarrow|r\rangle$, but those along $\mathbf{l}_\shortparallel$ is for the transition between $|r\rangle$ and another Rydberg state $|R\rangle$ via a low-lying intermediate state. Because $|r\rangle$ is near $|R\rangle$, the phase $\varphi_\shortparallel$ can be easily set in method II. The discussion about the phase above is applicable for both $\varphi_\shortparallel$ and $\varphi_\perp$ in method I and $\varphi_\perp$ in method II.

It is useful to briefly show why two methods are introduced. For brevity, we use ``one pulse'' when the laser fields along $\mathbf{l}_\perp$ and $\mathbf{l}_\shortparallel$ irradiate the system simultaneously for a certain duration. In method I, one pulse is used for the Rydberg excitation, and the nontarget atoms can have some residual population in the Rydberg state when the target atom is excited to the Rydberg state. After the second pulse, both the target and nontarget atoms return to the ground state. No phase shift occurs for the target atom because of the spin echo. In method II, two pulses are used for Rydberg excitation. The two pulses form an optical spin echo sequence for the nontarget atoms so that they have no population in the Rydberg state when the target atom is in the Rydberg state. Similarly, two similar pulses can pump the target atom back to its ground state. Because the target atom experiences no spin echo, it can accumulate a $\pi$ phase shift upon its state restoration. When there is Rydberg interaction between nontarget atoms, a microwave spin echo is used to reverse the sign of the Rydberg interaction, shown in Sec.~\ref{sec03D}. In Sec.~\ref{sec03E}, we study the issue of divergence of the laser beam. In Sec.~\ref{sec03F}, we study the application of the methods in the Rydberg blockade gate. We give a detailed comparison between the two methods in Sec.~\ref{sec03G}.

\subsection{Method I: one pulse for Rydberg excitation}\label{sec03A}

We first show Rydberg excitation of the target atom by one pulse. There are three stages to implement the optical spin echo, two optical pumping and one wait time. The wait can have a duration of $t_\mu=2\pi/\Delta$, during which a microwave field can be used to induce a transition between two different Rydberg states as discussed later; for Rydberg blockade gate, another atom near the target atom can be excited to the Rydberg state during the wait. 

In the first stage, laser fields along $\mathbf{l}_\perp$ and along $\mathbf{l}_\shortparallel$ induce the following transition
\begin{eqnarray}
\hat{H}_\perp^{(1)}&=& 
  \Omega_\perp e^{ it\Delta}|r\rangle_\perp\langle1|/2+\text{H.c.},\nonumber\\
\hat{H}_\shortparallel^{(1)}&=& 
  -\Omega_\shortparallel e^{ -it\Delta}|r\rangle_\shortparallel\langle1|/2+\text{H.c.} \label{HamiltonianPerp01}
\end{eqnarray}
during $t\in[0,~t_0)$, where $t_0=\pi/\Delta$ and the subscript $\alpha=\perp(\shortparallel)$ distinguishes notations for atoms in the two different light paths. For Gaussian beams, the values of  $\Omega_\alpha$ reach to their maximum at the target atom where $\Omega= \pi\Delta/4$. The above Hamiltonian is in a rotating frame with $\hat{\mathcal{R}}=\omega|r\rangle\langle r|$. To show the echo for the nontarget atoms, we use the rotating frame with $\hat{\mathcal{R}}_{\mp}\equiv\hat{\mathcal{R}}\mp \Delta|r\rangle\langle r|$ for the atoms along the two different light paths, and use $\psi~(\Psi)$ to denote the wavefunction in the interaction~(Schr\"{o}dinger) picture. The wavefunctions in the frame $\hat{\mathcal{R}}_{\mp}$ are  
\begin{eqnarray}
 | \psi(t)\rangle_\perp &=& e^{it\hat{\mathcal{R}}_{-} }  | \Psi(t)\rangle_\perp ,\nonumber\\ 
 | \psi(t)\rangle_\shortparallel &=& e^{it\hat{\mathcal{R}}_{+} }  | \Psi(t)\rangle_\shortparallel , \nonumber 
\end{eqnarray}
and the Hamiltonians are
\begin{eqnarray}
 \hat{\mathcal{H}}_\perp^{(1)} &=& \Delta|r\rangle_\perp\langle r| + \Omega_\perp(|r\rangle_\perp\langle 1| +\text{H.c.})/2, \nonumber\\ 
\hat{\mathcal{H}}_\shortparallel ^{(1)}&=& -\Delta|r\rangle_\shortparallel\langle r| - \Omega_\shortparallel(|r\rangle_\shortparallel\langle 1| +\text{H.c.})/2.   \label{HamiltonianPerp01-2}
\end{eqnarray}
The wavefunction evolves according to $e^{-i\hat{\mathcal{H}}_\alpha t} | \psi(0)\rangle_\alpha$, where $| \psi(0)\rangle_\alpha=|1\rangle_\alpha$ is the initial wavefunction. In Eqs.~(\ref{HamiltonianPerp01}) and~(\ref{HamiltonianPerp01-2}), the Rabi frequency $\Omega$ for the target atom is equal to the value of $\Omega_\perp$~(or $\Omega_\shortparallel$) at the foci of the laser fields. Because of the Gaussian profile of the fields, $\Omega$ is larger than $\Omega_\perp$ and $\Omega_\shortparallel$ of any nontarget atom. 
Equation~(\ref{HamiltonianPerp01}) means that the state of the target atom evolves according to the Hamiltonian
\begin{eqnarray}
  \hat{H} (t) &=&  i\Omega \sin(t\Delta) (|r\rangle\langle 1| -|1\rangle\langle r|) \label{Htarget01}
\end{eqnarray}
in the rotating frame of $\hat{\mathcal{R}}$, and one can show that when the initial state of the target atom is $|1\rangle$, its state becomes $|r\rangle$ at the end of the first pulse.

Second, a wait time of duration $t_\mu$ is inserted when nothing is done. If we would like to suppress the many-body effect as discussed later, a microwave transition can be used to transfer the state $|r\rangle$ to another Rydberg state $|r'\rangle$. There can be a phase difference between $\langle r|\psi(t_0)\rangle$ and $\langle r'|\psi(t_0+t_\mu)\rangle$ determined by the microwave fields, and the subsequent laser fields in the third stage should appropriately compensate this phase. But to show the essence of the optical spin echo, we assume nothing is done during the wait. The state of a nontarget atom evolves to
\begin{eqnarray}
 | \Psi(t_0+t_\mu)\rangle_\perp  &=&e^{-it_\mu\hat{\mathcal{R}} }e^{-it_0\hat{\mathcal{R}}_{-} } e^{-it_0 \hat{\mathcal{H}}_\perp^{(1)} } | \psi(0)\rangle_\perp  ,\nonumber\\
 | \Psi(t_0+t_\mu)\rangle_\shortparallel  &=&e^{-it_\mu\hat{\mathcal{R}} }e^{-it_0\hat{\mathcal{R}}_{+} } e^{-it_0 \hat{\mathcal{H}}_\shortparallel ^{(1)}} | \psi(0)\rangle_\shortparallel  ,
\end{eqnarray}
at the end of the wait.

Third, during $t\in[t_0+t_\mu,~2t_0+t_\mu)$, laser fields along both $\mathbf{l}_\perp$ and $\mathbf{l}_\shortparallel$ are sent for the following transitions,
\begin{eqnarray}
\hat{H}_\perp^{(2)}&=& 
  -\Omega_\perp e^{ -it\Delta}|r\rangle_\perp\langle1|/2+\text{H.c.},\nonumber\\
\hat{H}_\shortparallel^{(2)} &=& 
  \Omega_\shortparallel e^{ it\Delta}|r\rangle_\shortparallel\langle1|/2+\text{H.c.}, \label{HamiltonianPerp02}
\end{eqnarray}
so that the Hamiltonian for the target atom is still given by Eq.~(\ref{Htarget01}). When $t_\mu=2\pi/\Delta$, one can show that the state of the target atom is $| \Psi(t)\rangle =|1\rangle$ at the time $t=2t_0+t_\mu$. Now, we use the rotating frame with $\hat{\mathcal{R}}_{\pm}$ for the atoms along the two different light path so that the Hamiltonians become
\begin{eqnarray}
 \hat{\mathcal{H}}_\perp^{(2)} &=& -\Delta|r\rangle_\perp\langle r| - \Omega_\perp(|r\rangle_\perp\langle 1| +\text{H.c.}), \nonumber\\ 
\hat{\mathcal{H}}_\shortparallel ^{(2)}&=& \Delta|r\rangle_\shortparallel\langle r| + \Omega_\shortparallel(|r\rangle_\shortparallel\langle 1| +\text{H.c.}).  \label{HamiltonianPerp02rotate}
\end{eqnarray}
The state of the nontarget atoms becomes,
\begin{eqnarray}
  | \psi(2t_0+t_\mu)\rangle_\perp  &=&e^{-it_0 \hat{\mathcal{H}}_\perp^{(2)} } e^{i(t_0+t_\mu)\hat{\mathcal{R}}_{+} } e^{-it_\mu\hat{\mathcal{R}} }e^{-it_0\hat{\mathcal{R}}_{-} } \nonumber\\
  &&\times e^{-it_0 \hat{\mathcal{H}}_\perp^{(1)} } | \psi(0)\rangle_\perp  ,\nonumber\\
 | \psi(2t_0+t_\mu)\rangle_\shortparallel  &=&e^{-it_0 \hat{\mathcal{H}}_\shortparallel ^{(2)}} e^{i(t_0+t_\mu)\hat{\mathcal{R}}_{-} } e^{-it_\mu\hat{\mathcal{R}} }e^{-it_0\hat{\mathcal{R}}_{+} }\nonumber\\
  &&\times e^{-it_0 \hat{\mathcal{H}}_\shortparallel ^{(1)}} | \psi(0)\rangle_\shortparallel   ,\label{method01finalstate}
\end{eqnarray}
Because $e^{i(t_0+t_\mu)\hat{\mathcal{R}}_{+} } e^{-it_\mu\hat{\mathcal{R}} }e^{-it_0\hat{\mathcal{R}}_{-} }=e^{4i\pi |r\rangle\langle r|}$, it is equal to the identity since it operates either on $|r\rangle$ or $|1\rangle$; similarly, $e^{i(t_0+t_\mu)\hat{\mathcal{R}}_{-} } e^{-it_\mu\hat{\mathcal{R}} }e^{-it_0\hat{\mathcal{R}}_{+} }=\hat{1}$. Because $\hat{\mathcal{H}}_\alpha^{(2)} =- \hat{\mathcal{H}}_\alpha^{(1)} $, we have
\begin{eqnarray}
  | \psi(2t_0+t_\mu)\rangle_\perp  &=&e^{-it_0 \hat{\mathcal{H}}_\perp^{(2)} } e^{-it_0 \hat{\mathcal{H}}_\perp^{(1)} } | \psi(0)\rangle_\perp =| \psi(0)\rangle_\perp  ,\nonumber\\
  | \psi(2t_0+t_\mu)\rangle_\shortparallel  &=&e^{-it_0 \hat{\mathcal{H}}_\shortparallel ^{(2)}} e^{-it_0 \hat{\mathcal{H}}_\shortparallel ^{(1)}} | \psi(0)\rangle_\shortparallel  = | \psi(0)\rangle_\shortparallel   ,\nonumber\\
  \label{method01finalstate2}
\end{eqnarray}
 where $\alpha=\perp,\shortparallel$. So, the state of the nontarget atom is restored whatever the magnitude of $\Omega_\alpha$ is.
The value of $t_\mu=2\pi/\Delta$ is chosen so that Eq.~(\ref{method01finalstate}) can be simplified to Eq.~(\ref{method01finalstate2}). In fact, it can be an arbitrary value of $t_\mu'\in(0,~ t_\mu)$, but each set of the laser fields used during the third stage should have an extra phase $\varphi=t_\mu'\Delta$ so that $(\Omega_\perp,~\Omega_\shortparallel)$ becomes $(\Omega_\perp e^{i\varphi},~\Omega_\shortparallel e^{- i\varphi})$.

\begin{figure}
\includegraphics[width=3.2in]
{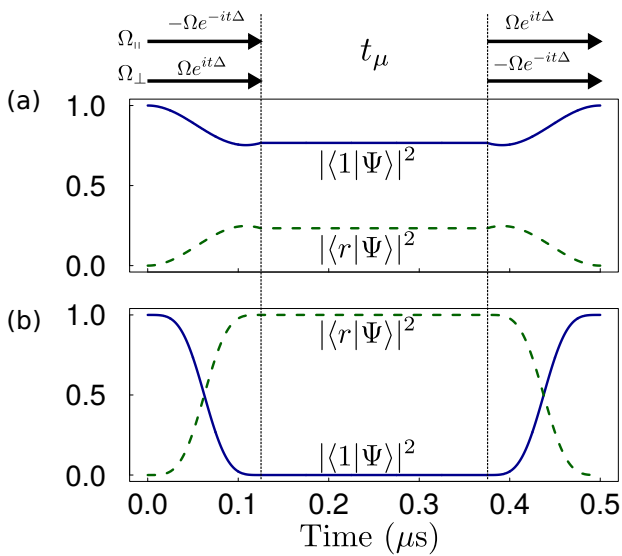}
\caption{Numerical result for the excitation and deexcitation of Rydberg state $|r\rangle$ in Method I. (a) and (b) show the time evolution of the population in $|1\rangle$~(solid curve) and $|r\rangle$~(dashed curve) for the nontarget atom and target atom, respectively, with $(\Delta,~\Omega)/2\pi=(4,~\pi)$~MHz, $t_\mu=2\pi/\Delta$, and the Rabi frequency for the nontarget atom is $\Omega_\perp=0.73\Omega$. The same state evolution occurs if $t_\mu=0$. \label{figure-method1} }
\end{figure}

As shown above, the merit of the method is that it can avoid bringing other atoms to Rydberg states upon the completion of the deexcitation of the target atom. The method works well when the light intensity is different at atoms along the light path. The focused Gaussian beam, even with a very small waist, can still have a Rayleigh length as large as $X=26~\mu$m~\cite{Wang2016}. At a spot away from the foci by $\mathfrak{l}_0$ for the beam propagating along $\mathbf{l}_\shortparallel$, the laser intensity at the beam axis is smaller than that at the foci by about $X^2/(X^2+\mathfrak{l}_0^2)$. This fact leads to different magnitudes of Rabi frequency at different nontarget atoms, and its irrationality indicates that it is impossible to simultaneously restore the states of all the atoms along the light path unless spin echo is used.

We continue to analyze the condition when the blockade interaction is included. If the lattice constant is $\mathcal{L}=10~\mu$m~\cite{Isenhower2010}, the values of $\Omega_\alpha=\Omega_\shortparallel$~(or $\Omega_\perp$) for a nontarget atom and $\Omega$ of the target atom are comparable because there are only two nontarget atoms in the light path along $\mathbf{l}_\shortparallel$, denoted as atom $A$ and atom $G$ in Fig.~\ref{figure10}(a). Atom $A$ is away from the foci by $\mathfrak{l}_0=\pm\sqrt6\mathcal{L}$, and thus its Rabi frequency $\Omega_\alpha$ is only about $0.73\Omega$ when $X=26~\mu$m. We suppose that the van der Waals interaction between Rydberg atoms is $V=C_6/\mathcal{L}^6$ and $V/\Omega\approx12$~[see text following Eq.~(\ref{OurGate}) for the reason of this choice]. Then, the blockade interaction between the target atom and the nontarget atom is only $V/216$. On the other hand, the Rydberg population in the nontarget atom (if it is initialized in the state $|1\rangle$) is tiny, shown above, and thus the influence of the blockade interaction for the excitation of the target atom is negligible. For the same reason, the Rydberg blockade does not perturb the spin-echo time evolution in the nontarget atom since $V/216\ll \Omega_\alpha$. So, the Rydberg addressing of one atom in a small 3D lattice with 45 sites in Fig.~\ref{figure10} is achievable. 

Using Eqs.~(\ref{HamiltonianPerp01}) and~(\ref{HamiltonianPerp02}) in the interaction picture with $\hat{\mathcal{R}}$, we numerically study the state evolution of the target atom and the nontarget atom with parameters $(\Delta,~\Omega)/2\pi=(4,~\pi)$~MHz, $t_\mu=2\pi/\Delta$ and, as an example, $\Omega_\perp=\Omega_\shortparallel=0.73\Omega$. Because the population evolution for the nontarget atom $\alpha=\perp$ is the same with that for $\alpha=\shortparallel$, we take $\alpha=\perp$ as an example and ignore the Rydberg blockade. The results are shown in Fig.~\ref{figure-method1}, where the population in $|r\rangle$ is 0.23 during $t\in[t_0,~t_0+t_\mu)$ for the nontarget atom when the target atom is in the Rydberg state. As a result of the optical spin echo, the state of the nontarget atom returns to the ground state at the end of the pulse sequence.

  One concern about the applicability of this method is that there is Rydberg-state decay in any irradiated nontarget atom during the pulse sequence, which can cause decoherence for many qubits. But the decay is negligible: if the state is $|\psi(0)\rangle=|1\rangle$ for a nontarget atom, the probability of decay for the atom is $T_{\text{de}}/\tau$, where $\tau$ is the lifetime of the Rydberg state, and 
 \begin{eqnarray}
  T_{\text{de}}&=& \int_0^{2t_0+t_\mu} |\langle r| \Psi(t)\rangle|^2 dt,
 \end{eqnarray}
 which is $0.093~\mu$s and $0.375~\mu$s for the processes in Figs.~\ref{figure-method1}(a) and~\ref{figure-method1}(b), respectively. For an $s$ or $d-$orbital state of rubidium with principal value of $100$, $\tau$ is about $320~\mu$s in a temperature of $300$~K~\cite{Beterov2009}, which gives a decay error of about $E_{\text{decay-n}}=2.9\times10^{-4}$ for the nontarget atom, and $1.2\times10^{-3}$ for the target atom; when the sequence in Fig.~\ref{figure-method1} is used for the pumping of the control qubit in the Rydberg blockade gate~\cite{PhysRevLett.85.2208}, the decay of one nontarget atom contributes an error $E_{\text{decay-n}}/2$ to the gate fidelity. Moreover, the state of a nontarget qubit can also be $|0\rangle$ that doesn't respond to the light irradiation. The farther the nontarget atom is from the foci, the smaller the value of $\Omega_\perp$ will be. With $\Omega_\perp=0.3\Omega$, we have $T_{\text{de}}=0.02~\mu$s which gives a decay error of about five times smaller than that in Fig.~\ref{figure-method1}(a). These analyses show that the Rydberg-state decay for the nontarget atoms is negligible.

\subsection{Method II: two-pulse excitation of Rydberg states}\label{sec03B}
For high fidelity quantum control, it is desirable to avoid Rydberg-state decay whenever possible. The scheme in Sec.~\ref{sec03A} has a wait time during which the nontarget atoms have some probability in the Rydberg state, shown in Fig.~\ref{figure-method1}(a). This leads to extra Rydberg-state decay that hampers the protocol. Besides, the target atom can not acquire any phase shift upon its state restoration because of the spin echo. In this section, we show a method that leaves no Rydberg population in the nontarget atom when the target atom is excited to the Rydberg state. Moreover, the target atom can have a $\pi$ phase shift when its state is restored. For brevity, we only show the Rydberg excitation of the target atom since its deexcitation is achieved with a similar process.

In the first pulse of duration $t_0=\pi/\Delta$, laser fields along $\mathbf{l}_\perp$ similar to that in Eq.~(\ref{HamiltonianPerp01}) are used for the following transition,
\begin{eqnarray}
\hat{H}_\perp^{(1)}&=& 
  \Omega_\perp e^{ it\Delta}|r\rangle_\perp\langle1|/2+\text{H.c.}, \label{method2-01}
\end{eqnarray}
and meanwhile laser fields along $\mathbf{l}_\shortparallel$ induce a two-photon transition between $|r\rangle$ and $|R\rangle$ via a low-lying intermediate state $|p\rangle$,
\begin{eqnarray}
\hat{H}_\shortparallel^{(1)}&=& 
  \Omega_\shortparallel e^{ -it\Delta}|R\rangle_\shortparallel\langle r|/2+\text{H.c.}, \label{method2-02}
\end{eqnarray}
where $|p\rangle$ should be a state higher than $|e\rangle$, and the fields used for $|r\rangle\rightarrow|p\rangle\rightarrow|R\rangle$ should be optical or infrared.

After the first pulse, a wait duration $t_\mu$ elapses as in Sec.~\ref{sec03A}, where microwave pumping can be used if necessary~\cite{Shi2018prapp2}. As shown in Sec.~\ref{sec03A}, the duration of the wait can be adjusted without altering the optical spin echo, and hence we can assume $t_\mu=0$ for brevity.

In the second pulse of duration $t_0$, laser fields along $\mathbf{l}_\perp$ are sent for the following transition,
\begin{eqnarray}
\hat{H}_\perp^{(2)}&=& 
  -\Omega_\perp e^{- it\Delta}|r\rangle_\perp\langle1|/2+\text{H.c.}, \label{method2-03}
\end{eqnarray}
while the laser fields along $\mathbf{l}_\shortparallel$ are used for
\begin{eqnarray}
\hat{H}_\shortparallel^{(2)}&=& 
  \Omega_\shortparallel e^{ it\Delta}|R\rangle_\shortparallel\langle r|/2+\text{H.c.}. \label{method2-04}
\end{eqnarray}
Whatever the magnitude of $\Omega_\perp$ is, the state of a nontarget atom irradiated by the laser fields along $\mathbf{l}_\perp$ can return to its initial state, as shown in Sec.~\ref{sec03A}. Because there is no population in $|r\rangle$ or $|R\rangle$ for any nontarget atom irradiated by the fields along $\mathbf{l}_\shortparallel$, nothing happens for them. Below, we study the state evolution for the target atom.

\begin{figure}
\includegraphics[width=3.2in]
{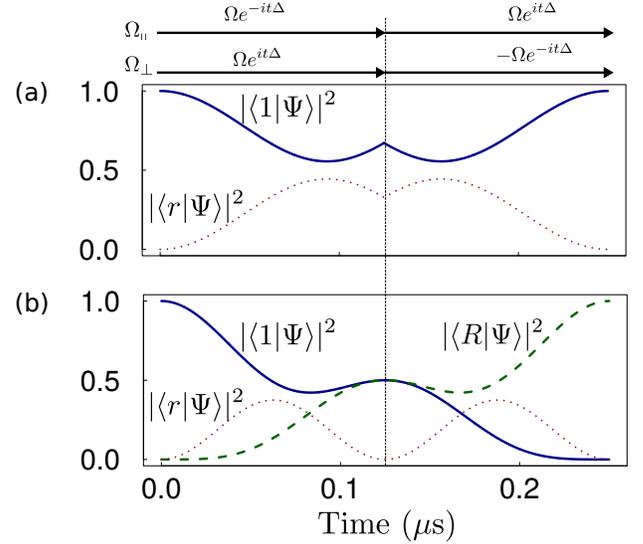}
\caption{Numerical result by Eqs.~(\ref{method2-01})-(\ref{method2-04}) for the excitation of the Rydberg state $|R\rangle$ in Method II. (a) shows the time evolution of the population in $|1\rangle$~(solid curve) and $|r\rangle$~(dotted curve) for the nontarget atom $E$ (or atom $F$, see Fig.~\ref{figure10}). The Rabi frequency for the nontarget atom $E$ is $\Omega_\perp=0.73\Omega$ as estimated in Sec.~\ref{sec03A}. (b) shows the time evolution of the population in $|1\rangle$~(solid curve), $|R\rangle$~(dashed curve), and $|r\rangle$~(dotted curve) for the target atom (labeled as atom $C$ in Fig.~\ref{figure10}), with $\Delta/2\pi=4$~MHz, $\Omega = 1.2247\Delta$, and $t_\mu=0$. In (b), the initial state is $|1\rangle$, and the final state is $i|R\rangle$ with an error smaller than $10^{-8}$. The same optical spin echo for the nontarget atom and the Rydberg excitation of the target atom are achieved with $t_\mu=2\pi/\Delta$.   \label{figure-method2} }
\end{figure}

Equations~(\ref{method2-01})-(\ref{method2-04}) are written in a rotating frame with 
\begin{eqnarray}
\hat{\mathcal{R}}&=& \omega|r\rangle\langle r|+\omega_R|R\rangle\langle R|,
\end{eqnarray}
where $\omega_R$ is the energy~(divided by $\hbar$) of the state $|R\rangle$. If we use a rotating frame with $\hat{\mathcal{R}}_\mp= \hat{\mathcal{R}}\mp\Delta|r\rangle\langle r|$ for the first~(second) pulse, the Hamiltonians of the target atom become
\begin{eqnarray}
 \hat{\mathcal{H}}^{(1)} &=& \Delta|r\rangle\langle r| + (\Omega_\shortparallel|R\rangle\langle r|+\Omega_\perp|r\rangle\langle 1| +\text{H.c.})/2,\nonumber\\
 \hat{\mathcal{H}}^{(2)} &=& -\Delta|r\rangle\langle r| + (\Omega_\shortparallel|R\rangle\langle r|-\Omega_\perp|r\rangle\langle 1| +\text{H.c.})/2 .  \nonumber\\\label{method2-06}
\end{eqnarray}
For $\Delta\gg \Omega_\perp,\Omega_\shortparallel$, an effective pumping emerges between $|1\rangle$ and $|R\rangle$ with a Rabi frequency of $\Omega_\perp\Omega_\shortparallel/(2\Delta)$ for both $\hat{\mathcal{H}}^{(1)}$ and $\hat{\mathcal{H}}^{(2)}$. But the time for a complete excitation of $|R\rangle$ is $2\pi\Delta/(\Omega_\perp\Omega_\shortparallel)$. If $N$ cycles of optical spin echo sequence are used, the condition of $\Delta/\sqrt{\Omega_\perp\Omega_\shortparallel}=\sqrt{N}$ leads to excitation of the target qubit. However, this is complicated. For an efficient quantum control, the case of $N=1$ is best. We numerically found that with $\Omega_\perp=\Omega_\shortparallel=1.2247\Delta$, a complete excitation of the ground state $|1\rangle$ to the Rydberg state $i|R\rangle$ is achieved with one spin-echo cycle, shown in Fig.~\ref{figure-method2}; the residual population in $|1\rangle$ and $|r\rangle$ is $1.9(5.6)\times10^{-9}$ at the time of $t=2t_0$ in Fig.~\ref{figure-method2}(b). The time for the nontarget atom~[atom $E$ of Fig.~\ref{figure10}(a)] to be in the Rydberg state is $ T_{\text{de}}=0.07~\mu$s in Fig.~\ref{figure-method2}(a), which means that the decay probability for the nontarget atom is $ T_{\text{de}}/\tau=2\times10^{-4}$ if $\tau=320~\mu$s, i.e., negligible.

In Fig.~\ref{figure-method2}(a), the population dynamics for atom $E$ of Fig.~\ref{figure10} was shown. As estimated in the third paragraph counted backward from the end of Sec.~\ref{sec03A}, if the Rabi frequency in Eq.~(\ref{method2-01}) for atom $C$ is $\Omega$, then atom $E$ is pumped according to Eq.~(\ref{method2-01}) with a Rabi frequency $0.73\Omega$, which was used in Fig.~\ref{figure-method2}(a). The reason to ignore the population of $|R\rangle$ in Fig.~\ref{figure-method2}(a) lies in that the field amplitude for the transition $|r\rangle\leftrightarrow|R\rangle$ at atom $E$ is negligible:  the distance from $E$ to the beam axis of the field propagating along $\mathbf{l}_\shortparallel$ is $r_{\perp,E}\approx2.3\mathcal{L}$, while atom $B$ is only away from this beam axis by about $r_{\perp,B}\approx0.46\mathcal{L}$. The field drops off by the law of $\propto e^{-r_{\perp}^2/ w^2}$ at a point away from the beam axis by $r_{\perp}$, where $w$ is the beam radius. As detailed later below Eq.~(\ref{Sec04b03Eq01}), we can suppose that the field amplitude at atom $B$ is no more than $1/e$ of that at the beam axis, i.e., $e^{-r_{\perp,B}^2/ w^2}\leq 1/e$. Because $r_{\perp,E}^2/r_{\perp,B}^2\approx25$, the field amplitude for the transition $|r\rangle\leftrightarrow|R\rangle$ at atom $E$ can be ignored. Similarly, we can ignore any pumping of atom A (or atom G) by the field propagating along $\mathbf{l}_\perp$. In Fig.~\ref{figure10}(a), there are other atoms very near to the beam axes, such as atom $D$. But it is away from the beam axis of the fields along either $\mathbf{l}_\perp$ or $\mathbf{l}_\shortparallel$ by about $0.9\mathcal{L}\approx 2r_{\perp,B}$, which means that the field amplitude for either beam at atom $D$ is smaller than $1/e^4$ of that at the beam axis if conditions like $e^{-r_{\perp,B}^2/ w^2}\leq 1/e$ apply to both beams. This means that only the target atom, i.e., atom $C$ experiences pumping by the two laser beams.

In Fig.~\ref{figure-method2}(b), the final phase $\pi/2$ for the state $|R\rangle$ arises in response to the sequential pumping with $(\hat{H}_\perp^{(1)},\hat{H}_\shortparallel^{(1)})$ and $(\hat{H}_\perp^{(2)},\hat{H}_\shortparallel^{(2)})$ defined in Eqs.~(\ref{method2-01})-(\ref{method2-04}); the same pulse sequence as in Fig.~\ref{figure-method2} can deexcite the target atom from $i|R\rangle$ to $-|1\rangle$ which can lead to the conditional $\pi$ phase if method II is used for the Rydberg blockade gate~\cite{PhysRevLett.85.2208}. Note that if only $(\hat{H}_\perp^{(1)},\hat{H}_\shortparallel^{(1)})$ is used for Rydberg excitation with a time of $2t_0$, the final state of the target atom is $-|R\rangle$, but a subsequent deexcitation by this sequence can not lead to a conditional $\pi$ phase for entanglement generation. 

Because the energy difference between $|r\rangle$ and $|R\rangle$ is smaller than that between $|r(1)\rangle$ and $|p\rangle$ by several orders of magnitude, the wavevector for $\Omega_\shortparallel$ is negligible when the field for $|r\rangle\rightarrow|p'\rangle$ and the field for $|p'\rangle\rightarrow|R\rangle$ copropagate, where $|p(p')\rangle$ is the intermediate state for $|r\rangle\leftrightarrow|1(R)\rangle$. Thus, it is easy to set the phase for $\Omega_\shortparallel$ although that for $\Omega_\perp$ is still an issue. In this sense, the position distribution of the qubit will induce error only through the phase fluctuation of $\Omega_\perp$, and the fidelity for method II can be much larger than that of method I. For more detail about this issue, see the third paragraph at the beginning of Sec.~\ref{sec03}.

\subsection{Rydberg addressing in a relatively large 3D lattice}\label{sec03D}
For a nontarget atom away from the foci of the Gaussian beam, the relevant Rydberg blockade can be comparable to the Rabi frequency $\Omega_\alpha$ for the nontarget atom in the light path along $\mathbf{l}_\alpha$, where $\alpha=\perp,\shortparallel$. Then, we should include the blockade in the analysis about the state evolution of the nontarget atom. First of all, because any nontarget atom lying away from the beam axis is pumped by a very small field, its population in Rydberg states is negligible and for a first approximation it does not induce dynamical phase in the target atom. Second, the blockade between two atoms along the beam axis is $V/216$ as shown in Sec.~\ref{sec03A}, so the blockade interaction between the target atom and a nontarget atom can be neglected. Then, we can separate the atoms influenced by the lasers into two groups: the target atom, and all the nontarget atoms. These two groups do not disturb each other. But there is many-body physics in the second group: for two nontarget atoms away from the foci of the laser beams by $\mathfrak{l}_0\gg X$, $\Omega_\alpha$ can be comparable to the interaction $V/216$ between them because $\Omega_\alpha/\Omega\propto X/\sqrt{X^2+\mathfrak{l}_0^2}$. So, the single-particle optical spin echo in Secs.~\ref{sec03A} and~\ref{sec03B} is not sufficient to restore the state of the nontarget atoms. 

\subsubsection{Microwave spin echo for method I}\label{sec03D1}
A microwave spin-echo sequence can eliminate the many-body imprinted Rydberg excitation in the nontarget atoms. As an example, we take method I shown in Sec.~\ref{sec03A} to explain this. The required modification is that between the two stages of time evolution shown in Hamiltonian~(\ref{HamiltonianPerp01}) and~(\ref{HamiltonianPerp02}), we should add a two-photon microwave transition between $|r\rangle$ and another Rydberg state $|r'\rangle$
\begin{eqnarray}
\hat{H}_\mu&=& 
  i\Omega_\mu |r'\rangle\langle r|/2+\text{H.c.}, \label{micr01}
\end{eqnarray}
where the microwave field covers all atoms in the system, but will only influence states that are initially in $|r\rangle$; one can assume that the strength of the microwave field is the same for all atoms in the system. The Rabi frequency of this microwave field should be much larger than the blockade interactions between any two atoms in the problem, and the signs of the blockade interactions of $|rr\rangle$ and $|r'r'\rangle$ should be opposite. This latter requirement is achievable if $|r\rangle$ and $|r'\rangle$ are $s$- and $d$-orbital states of $^{87}$Rb, respectively~\cite{Walker2008}. If the ratio between the $C_6$ coefficients of $|r'r'\rangle$ and $|rr\rangle$ is $\varkappa$, the amplitude and detuning of the laser fields used in the latter stage of Eq.~(\ref{HamiltonianPerp02}) should be $\varkappa \Omega$ and $\varkappa \Delta$, respectively. Then, the duration of the second pulse in the microwave spin echo becomes $t_1=\pi/|\varkappa \Delta|=t_0/|\varkappa|$. Because $\varkappa$ can differ from $-1$, it is necessary to either add appropriate phases in the laser fields, or add extra wait time. To explain this, we take a single nontarget atom as an example. Suppose the energy of the state $|r'\rangle$ is $\hbar\omega'$, then the rotating frame with it is given by $\hat{\mathcal{R}}'=\hat{\mathcal{R}}+\omega'|r'\rangle\langle r'|$, and the operators $\hat{\mathcal{R}}_\pm$ in Secs.~\ref{sec03A} and~\ref{sec03B} change to $\hat{\mathscr{R}}_\pm= (\omega'\pm|\varkappa|\Delta)|r'\rangle\langle r'|$, which are applicable after the microwave pulse. If the microwave field lasts for a duration of $t_\mu$, an extra wait time $t_{\text{w}}$ changes Eq.~(\ref{method01finalstate}) to
\begin{eqnarray}
  | \psi(t_{\text{f}})\rangle_\perp  &=&e^{-it_1 \hat{\mathscr{H}}_\perp^{(2)} } e^{i(t_{\text{w}} +t_\mu+ t_0)\hat{\mathscr{R}}_{+} }e^{-i(t_{\text{w}} +t_\mu+ t_0)\hat{\mathcal{R}}' }  \nonumber\\
  &&\times e^{-it_\mu\hat{H}_\mu } e^{it_0\hat{\mathcal{R}}' }e^{-it_0\hat{\mathcal{R}}_{-} } e^{-it_0 \hat{\mathcal{H}}_\perp^{(1)} } | \psi(0)\rangle_\perp  ,\nonumber\\
 | \psi(t_{\text{f}})\rangle_\shortparallel  &=&e^{-it_1 \hat{\mathscr{H}}_\shortparallel ^{(2)}} e^{i(t_{\text{w}} +t_\mu+ t_0)\hat{\mathscr{R}}_{-} }e^{-i(t_{\text{w}} +t_\mu+ t_0)\hat{\mathcal{R}}' }    \nonumber\\
  &&\times e^{-it_\mu\hat{H}_\mu } e^{it_0\hat{\mathcal{R}}' }e^{-it_0\hat{\mathcal{R}}_{+} } e^{-it_0 \hat{\mathcal{H}}_\shortparallel ^{(1)}} | \psi(0)\rangle_\shortparallel   , \label{sec3Dequa01}
\end{eqnarray}
where $t_{\text{f}} = t_1+t_{\text{w}} +t_\mu+t_0$ is the time at the end of the sequence, and 
\begin{eqnarray}
\hat{\mathscr{H}}_\perp^{(2)}&=& 
  -|\varkappa|\Omega_\perp e^{ -it|\varkappa|\Delta}|r'\rangle_\perp\langle1|/2+\text{H.c.},\nonumber\\
\hat{\mathscr{H}}_\shortparallel^{(2)} &=& 
 |\varkappa| \Omega_\shortparallel e^{ it|\varkappa|\Delta}|r'\rangle_\shortparallel\langle1|/2+\text{H.c.}. 
\end{eqnarray}
Compared to Eq.~(\ref{method01finalstate}), the different magnitudes of the detuning of the laser fields before and after the wait in Eq.~(\ref{sec3Dequa01}) requires appropriate wait time to validate the microwave spin echo. To show this, one can verify that with $\Omega_\mu t_\mu=\pi$, the operator $e^{-it_\mu\hat{H}_\mu }$ is equivalent to $(|r'\rangle\langle r|-$H.c.) when acting on a state $|r(r')\rangle$, and is $\hat{1}$ otherwise. Thus the microwave spin echo in the two equations in Eq.~(\ref{sec3Dequa01}) requires $|\varkappa|\Delta(t_{\text{w}} +t_\mu+ t_0)+t_0\Delta=2n\pi$, where $n$ is an integer, which gives
\begin{eqnarray}
t_{\text{w}} +t_\mu &=& \frac{2n\pi}{|\varkappa|\Delta} - t_0(1+1/|\varkappa|).
\end{eqnarray}
For fast quantum control, the minimal positive integer $n$ leads to the smallest $t_{\text{w}} +t_\mu$. When the above condition is satisfied, the many-body Hamiltonian $\hat{H}'$ during $t\in[t_{\text{w}} +t_\mu+ t_0,~t_{\text{w}} +t_\mu+ t_0 +t_1)$ is exactly $\varkappa$ times the Hamiltonian during $t\in[0,~t_0)$; the only modification is that the involved Rydberg state is different. As a consequence, the many-body system of the nontarget atoms will have their states restored even though they experience many-body state evolution; in fact, even if there is interaction between the target and nontarget qubit, the microwave field also changes the state of the target atom, and hence the many-body effect can still be removed. For a numerical simulation, see the appendix of Ref.~\cite{Shi2018prapp2} where the duration of $t_{\text{w}} +t_\mu$ is ignored supposing the spin-echo condition is satisfied.

\subsubsection{Microwave spin echo for method II}  
The process above is applicable to method II, too. Because there is no nontarget atom in the light path along $\mathbf{l}_\shortparallel$, we focus on the nontarget atoms in the light path along $\mathbf{l}_\perp$. As in Sec.~\ref{sec03D1}, a microwave field pumps $|r\rangle$ to $|r'\rangle$ so as to reverse the sign of the Rydberg interaction. Because the microwave field influences all atoms, the state component $|r\rangle$ of the target qubit also changes likewise; but the target atom can have some population in $|R\rangle$, too. To completely remove the many-body effect in the system, we should reverse the sign of interaction in $|r'R\rangle$, too. This can be realized by using a superposition of two Rydberg states~\cite{ShiJPB2016,Shi2017pra,Shi2018a} as $|R\rangle$, so that its interaction with a nontarget atom can also change by a ratio of $\varkappa$. For more detail about this method, see Ref.~\cite{Shi2017pra}.

\begin{figure}
\includegraphics[width=3.2in]
{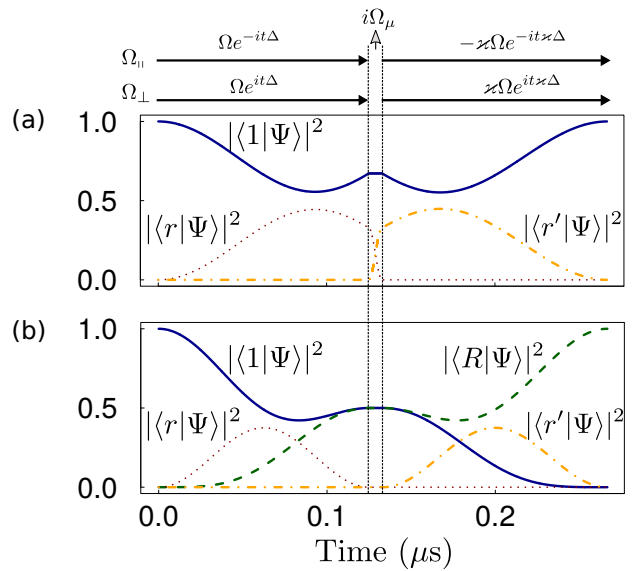}
\caption{Numerical result for the excitation of the Rydberg state $|R\rangle$ in Method II with a microwave field $\pi$-pulse between the two optical pulses. (a) shows the time evolution of the population in $|1\rangle$~(solid curve), $|r\rangle$~(dotted curve), and $|r'\rangle$~(dash-dotted curve) for the nontarget atom $E$ (or atom $F$). (b) shows the time evolution of the population in $|R\rangle$~(dashed curve), $|1\rangle$, $|r\rangle$, and $|r'\rangle$ for the target atom. Before the microwave pulse, $\Delta/2\pi=4$~MHz and $\Omega = 1.2247\Delta$; after the microwave pulse, the detuning and Rabi frequency become $|\varkappa|\Delta$ and $|\varkappa|\Omega$ with $\varkappa=-0.936$~(see text), and the microwave field lasts for a $\pi$-pulse duration of $t_\mu=(1/|\varkappa|-1)\pi/\Delta$ with $\Omega_\mu=\pi/t_\mu$. The Rabi frequency for the nontarget atom is $\Omega_\perp=0.73\Omega$ in (a).   \label{figure-method2mu} }
\end{figure}

In Fig.~\ref{figure-method2mu}, we show the microwave spin echo with one atom by using $\varkappa=-52.6/56.2$, which corresponds to the $100d(s)$ state of rubidium~\cite{Shi2018prapp2}. Figure~\ref{figure-method2mu} shows that the nontarget atom irradiated by the light along $\mathbf{l}_\perp$ indeed returns to the ground state, and the target atom is excited to the Rydberg state $|R\rangle$. The final phase of the Rydberg state is $\pi/2$; when the same spin-echo sequence in Fig.~\ref{figure-method2mu} is used for deexcitation, the final state is $-|1\rangle$ for the target atom.

\subsection{Limitation from divergence of laser beam}\label{sec03E}  
The above method assumes that at most one nontarget atom is irradiated near $\mathfrak{l}_0$ for each laser beam. If more than one nontarget atoms are irradiated near $\mathfrak{l}_0$, the Rydberg interactions between the nontarget atom at the axis of the beam and the atom off the axis can not have a ratio of $\varkappa$ in the two pulses of the optical spin echo. This is because the interaction between $s$-orbital Rydberg atoms is isotropic, but that between $d$-orbital atoms is not~\cite{Walker2008}. Even the Rydberg interaction along the the light path can change by a factor of $\varkappa$ after the microwave irradiation, the interaction between two atoms not along the light path can not.

We study condition that approximately validates the microwave spin echo in Sec.~\ref{sec03D}. We first consider two nontarget atoms shown in Fig.~\ref{figure10}(a), labeled as $A$ and $B$, where $B$ is not at the beam axis, but $A$ is, and thus atom $A$ is irradiated. Because the field amplitude decays rapidly away from the beam axis, atom $B$ is irradiated with a small field. For the configuration in Fig.~\ref{figure10}, the angle between $AC$ and $BC~(AB)$ is only $0.3~(0.4)$~radian, so that the interaction between them is approximately isotropic~\cite{Walker2008}, and if no atom that is even farther to the beam axis than B is irradiated, the many-body spin echo is valid. 

That no nontarget atoms farther than the atom B are irradiated is determined by the lattice constant and the laser beam. To study the decay of the light intensity away from the foci, we use $\mathfrak{l}_0$ as the longitudinal coordinate, and $r_\perp$ as the radial coordinate. The foci is at the atom $C$, which is the atom at the beam axis nearest to $A$. The coordinate of $B$ is $(\mathfrak{l}_0, r_{\perp0})$. From the geometry of the lattice, one can locate another atom $B'$ at $(\mathfrak{l}_1, 2r_{\perp0})$, where $\mathfrak{l}_1\approx\mathfrak{l}_0$. The beam radius of the laser field near $(\mathfrak{l}_1,0)$ is given by
   \begin{eqnarray}
     w(\mathfrak{l}_1) &=& w(0) \sqrt{X^2+\mathfrak{l}_1^2}/X\nonumber\\
     &=& \lambda\sqrt{\mathfrak{l}_1^2/w^2(0)+ \pi^2w^2(0)/\lambda^2}/\pi \label{Sec04b03Eq01}
\end{eqnarray} 
   where $w(0)$ is the beam radius at the beam waist and $\lambda$ is the wavelength of the laser field. The electric field amplitude of the laser is $\propto e^{-r_\perp^2/w^2(\mathfrak{l}_1) }$, which means that the electric field is $e^{-r_{\perp0}^2/w^2(\mathfrak{l}_1) }\mathcal{E}$~(or $e^{-4r_{\perp0}^2/w^2(\mathfrak{l}_1) }\mathcal{E}$) for the atom $B$~(or $B'$), where $\mathcal{E}$ is the amplitude at $(\mathfrak{l}_1, 0)$. If it is reasonable to neglect atomic excitation when the field amplitude of the laser light falls to $e^{-4}$ of that at the beam axis, $r_{\perp0}$ should be smaller than $w(0)$ so that $B'$ is not irradiated. From Eq.~(\ref{Sec04b03Eq01}), one finds that $ w(\mathfrak{l}_0) \geq \sqrt{2\lambda\mathfrak{l}_0/\pi }$. For Fig.~\ref{figure10}(a), one can show that $r_{\perp0}=0.46\mathcal{L}$. So, it is necessary to have $\mathfrak{l}_0<0.33\mathcal{L}^2/\lambda$ if we want the field at $B'$ to be smaller than $\mathcal{E}/e^4$. But for the sake of convenience in experiments, it is useful to let the two laser fields propagate only along the two directions, $[12\overline{1}]$ and $[\overline{1}21]$, so that by only sweeping the laser fields parallel one can address any site of the lattice. Then, we need to consider that the lasers can be focused at an atom that is at any edge of the lattice so as to derive the permissible largest lattice; one can find that the lattice size should be smaller than $(1+\mathcal{N})\times (1+2\mathcal{N})\times(1+ \mathcal{N})$, where $ \mathcal{N}$ is the largest integer that is smaller than $0.13\mathcal{L}/\lambda$. This means that only a lattice with a large enough lattice constant is applicable by our method. If small Rydberg blockade is needed, relatively large lattice constants like $16.5~\mu$m are required~\cite{Shi2019}; in such a case and with $\lambda=0.78~\mu$m, a lattice of the size $3\times 5\times 3$ in Fig.~\ref{figure10} can allow arbitrary single-site addressing by our method. For $\lambda\leq0.78~\mu$m in the case of rubidium, $0.13\mathcal{L}/\lambda\approx1$ if $\mathcal{L}=6~\mu$m, which means that a $2\times 3\times 2$ lattice can be addressed only if the lattice constant is as large as $6~\mu$m. This limitation from the beam divergence can be tackled by sending laser fields along various directions other than the two shown in Fig.~\ref{figure10} so that a larger lattice can be used for this purpose although it involves more experimental complexity. 

A possible solution to the problem of system size is to trap two types of atoms in one lattice. In Fig.~9 of Ref.~\cite{Beterov2015}, a two-dimensional lattice trapping both cesium and rubidium is studied. If such a lattice is extended to 3D in a configuration so that the atoms nearest to a rubidium are all cesium atoms, it should be possible to address the rubidium atoms with laser fields that propagate through cesium atoms. This method may allow more qubits to be trapped. However, it is an open question whether such a 3D lattice can be prepared.

\subsection{Application in Rydberg blockade gate}\label{sec03F} 
Till now, we have assumed that there is no Rydberg atom near the target atom before exciting it to Rydberg states. In the context of Rydberg blockade gate~\cite{PhysRevLett.85.2208}, if there is a Rydberg atom near the target qubit, it is called the control qubit. Before proceeding, we note that the microwave field used in the spin-echo sequence will influence both the control and target qubits if the same Rydberg state is used for the two qubits, and thus can transfer the Rydberg state of the control qubit. Then, the interaction between the control qubit and any nontarget atoms reverses the sign, validating the required spin-echo condition.

Compared to method I, method II is more useful when the control qubit is already in the Rydberg state. We assume that the control qubit is away from the target qubit by one lattice constant $\mathcal{L}$, and take the configuration in Fig.~\ref{figure10}(a) for analysis. For a concrete discussion, we assume that the atom labeled by $D$ in Fig.~\ref{figure10}(a) is the control qubit, which is already in the Rydberg state $|R_{1}\rangle$, and we are to excite the target atom $C$ to the Rydberg state $|R_{2}\rangle$ to prepare entanglement through the blockade interaction $V$ of the state $|R_1R_2\rangle$. Now atom $D$ is very near to the laser beams along both $\mathbf{l}_\shortparallel$ and $\mathbf{l}_\perp$, and one can choose different Rydberg states $|R_{1}\rangle$ and $|R_{2}\rangle$ when exciting the control and target qubit, respectively. When $|R_1\rangle$ and $|R_2\rangle$ have a GHz energy separation, the light field used for the target will not alter the state of the control qubit. If microwave spin echo is used in this case, three sets of microwave fields should be employed, one to transfer $|r\rangle$ to $|r'\rangle$~(as in Fig.~\ref{figure-method2mu}), and the other two for $|R_{1(2)}\rangle\rightarrow |R_{1(2)}'\rangle$. This can still realize the microwave spin echo when the interactions of the states $(|R_1r\rangle, |R_1R_2\rangle)$ and those of $(|R_1'r'\rangle, |R_1'R_2'\rangle)$ change by a common ratio $\varkappa$; to realize this, superposition states can be used as $|r'\rangle$ and $|R_{1(2)}'\rangle$~\cite{ShiJPB2016,Shi2017pra,Shi2018a}. But for method I, nontarget atoms along the laser fields along both $\mathbf{l}_\shortparallel$ and $\mathbf{l}_\perp$ can be excited to the Rydberg state. Then, because atom $B$ and atom $D$ have a distance of $\sqrt{3}\mathcal{L}$, the interaction between them is $V/3^3$, which can be comparable to the Rabi frequency and hampers the gate operation. This means that it is necessary to avoid exciting atom $B$ if method I is used, which puts a significant limit to the system size. But if method II is used, the light field along $\mathbf{l}_\shortparallel$ will not excite any Rydberg state, thus atom $B$ is not a problem. This means that both method I and method II can be used for the excitation of the control qubit, but method II is more useful for the $2\pi$ pulse for the target qubit in the Rydberg blockade gate~\cite{PhysRevLett.85.2208}.

\subsection{Comparison between method I and method II}\label{sec03G} 
The two methods shown in Secs.~\ref{sec03A} and~\ref{sec03B} have their own advantages and shortcomings, and can be used for different purposes.

First, both methods will leave the state of any nontarget atoms intact when the state of the target atom is restored to the ground state. However, method I only needs one pulse for exciting the target atom to the Rydberg state, while method II needs two.

Second, although the excitation and deexcitation of the target atom form an optical spin-echo cycle in method I, the excitation pulse alone does not form a spin echo. So, the nontarget atoms can have some residual population in Rydberg state when the target atom is pumped to the Rydberg state. For method II, Rydberg excitation of the target atom is by two sets of lasers: those along $\mathbf{l}_\shortparallel$ form a spin-echo cycle, but those along $\mathbf{l}_\perp$ do not. Although the field along $\mathbf{l}_\perp$ do not form a spin echo sequence, it does not excite any nontarget atoms, shown in Secs.~\ref{sec03B}. So, no nontarget atom will have any population in the Rydberg state when the target atom is completely excited to Rydberg states in method II.

Third, the state of the target atom will not pick any phase upon the completion of its state restoration in method I. For method II, the target atom does not experience any spin echo, and a $\pi$ phase is imprinted when its state is restored to the ground state. This means that for the Rydberg gate in Ref.~\cite{PhysRevLett.85.2208}, method I can be used for the control qubit, but can not be used for the target qubit, while method II can be used for both the control and target qubit.

    \section{High-fidelity quantum gates}\label{sec04b}
Another application of ORIR is in achieving high fidelity in a Rydberg quantum gate~\cite{PhysRevLett.85.2208,Saffman2010,Saffman2016,Weiss2017}. Although it is theoretically possible to achieve high fidelity beyond $0.999$ in an entangling quantum gate by Rydberg interactions~\cite{Goerz2014,Theis2016,Shi2017,Petrosyan2017,Shi2018prapp2,Shi2019}, the experimental implementation is difficult~\cite{Wilk2010,Isenhower2010,Zhang2010,Maller2015,Jau2015,Zeng2017,Picken2018}, with best fidelity below $0.98$~\cite{Graham2019,Levine2019}. This is partly due to a number of technical issues~\cite{DeLeseleuc2018}. But even if these issues are removed with improved technology~\cite{Levine2018,Graham2019,Levine2019}, the fidelity of the best-known Rydberg quantum gate, i.e., the Rydberg blockade gate~\cite{PhysRevLett.85.2208}, is limited by the blockade error~\cite{Zhang2012}. Below, we show that ORIR can effectively suppress this blockade error.

The Rydberg blockade gate maps the input states from $\{|00\rangle,|01\rangle,|10\rangle,|11\rangle \}$ to $\{|00\rangle,-|01\rangle,-|10\rangle,-|11\rangle \}$, and is realized in three steps: apply a $\pi$ pulse to the control qubit; apply a $2\pi$ pulse to the target; apply another $\pi$ pulse to the control qubit~\cite{PhysRevLett.85.2208}. These three steps induce the following state evolution for the input state $|11\rangle$:
\begin{eqnarray}
|11\rangle\xrightarrow[\text{For control}]{\pi}-i|r1\rangle\xrightarrow[\text{For target}]{2\pi}-i|r1\rangle\xrightarrow[\text{For control}]{\pi}-|11\rangle,\nonumber
\end{eqnarray}
where $|r\rangle$ is a Rydberg state. The second step in the above equation is perfect if the Rydberg blockade $V$ of the state $|rr\rangle$ is infinitely large compared to the Rabi frequency $\Omega$ of the atom-field interaction; for any finite $V$, it attains a blockade error on the order of $\Omega^2/V^2$. There is also a phase error, but it can be removed effectively by adding a phase difference in the laser field at the middle of the second step of the gate. When a global phase shift to the target qubit is inserted, the phase error is removed~\cite{Zhang2012}, but the blockade error sets a fundamental limit to the achievable fidelity of the gate. For this reason, there are many theoretical proposals to suppress the blockade error~\cite{Goerz2014,Theis2016,Shi2017,Petrosyan2017,Shi2018prapp2,Shi2019,Shen2019,Yu2019,Levine2019}.

We show that the blockade error can be suppressed by replacing the usual Rabi frequency $\Omega$ with a time-dependent Rabi frequency $i\Omega\sin(\Delta t)$ formed by ORIR. In practice, this is done via replacing a resonant laser of power $P$ by two laser beams of opposite phases, one with detuning $\Delta$ and power $P/4$, and the other with detuning $-\Delta$ and power $P/4$. This results in an effective Rabi frequency $i\Omega\sin(\Delta t)$. In contrast to the single-site Rydberg addressing studied in Sec.~\ref{sec03}, here the fields with detuning $\Delta$ and those with detuning $-\Delta$ can propagate in one direction, and hence it is easier to set the relative phase of the laser fields to realize Eq.~(\ref{sec02equation01}). In this case, the Hamiltonian for the input state $|11\rangle$ during the second step of the traditional protocol
\begin{eqnarray}
  \hat{H} &=&
\frac{1}{2}  \left(\begin{array}{cc}
   2V& \Omega\\
    \Omega &0
    \end{array}
  \right) \label{TraditionalGate}
\end{eqnarray}
is replaced by
\begin{eqnarray}
  \hat{H} &=&
  \left(\begin{array}{cc}
    V& i\Omega\sin(\Delta t)/2 \\
    -i\Omega\sin(\Delta t)/2 &0
    \end{array}
  \right) \label{OurGate}
\end{eqnarray}
in our scheme. The basis is $\{|rr\rangle,~|r1\rangle\}$ in the two matrices of Eqs.~(\ref{TraditionalGate}) and~(\ref{OurGate}). To induce the transform $| 01\rangle\rightarrow-| 01\rangle$ for the input state $| 01\rangle$, one can choose $\Omega/\Delta= \pi$ and a duration $t=\pi/\Delta$ for the second step, as can be derived in a similar method used for the derivation of Eq.~(\ref{CgCe}). Here, the condition of $\Omega/\Delta= \pi$ and $t=\pi/\Delta$ is not compatible with an optical spin echo sequence because it does not allow the transform from Eq.~(\ref{method01finalstate}) to Eq.~(\ref{method01finalstate2}), which is in contrast to the case studied in Sec.~\ref{sec03A}. This means that to use the method in this section for addressing one atom in 3D lattice, only atoms near the very edge of the atomic array can be excited so that no nontarget atoms will be excited.

To analyze the blockade errors for the two schemes in Eqs.~(\ref{TraditionalGate}) and~(\ref{OurGate}), we take the setup in~\cite{Levine2018} to estimate the system parameters. In~\cite{Levine2018}, its Figure~1 shows that the laser fields for the optical trap propagate~(along $\mathbf{z}$) perpendicular to the quantization axis~(labeled as $\mathbf{x}$) of the atom. The supplemental material of~\cite{Levine2018} shows that the r.m.s. fluctuation of the qubit spacing along $\mathbf{x}$ is $\varsigma_x=0.2~\mu$m. If the position fluctuation along $\mathbf{y}$ or $\mathbf{z}$ is about 10 times of that along $\mathbf{x}$ as in~\cite{Isenhower2010}, we can assume $\varsigma_y=\varsigma_z=2~\mu$m. The two-qubit spacing is $L=5.7~\mu$m in~\cite{Levine2018}, which means that the actual blockade $V$ can be very different from the chosen value $V_0$: fluctuation along $\mathbf{x}$ can lead to $V/V_0=(1\pm 2\varsigma_x/L)^{-6}~(\approx0.67,1.55)$, while that along $\mathbf{z}$ can lead to $V/V_0=(1+4\varsigma_z^2/L^2)^{-3}~(\approx0.3)$. For a conserved estimation, we consider $(V-V_0)/V_0\in[-0.25,~0.25]$. The entanglement between the ground and Rydberg states in~\cite{Levine2018} used $(\Omega,~V_0)/2\pi=(2,~30)$~MHz, while the actual Rabi frequency for the entanglement was $\sqrt2\Omega$ due to the many-body enhancement. Then, we take $V_0/\Omega=12$ as an example to study the rotation error of a Rydberg blockade process.
\begin{figure}
\includegraphics[width=3.2in]
{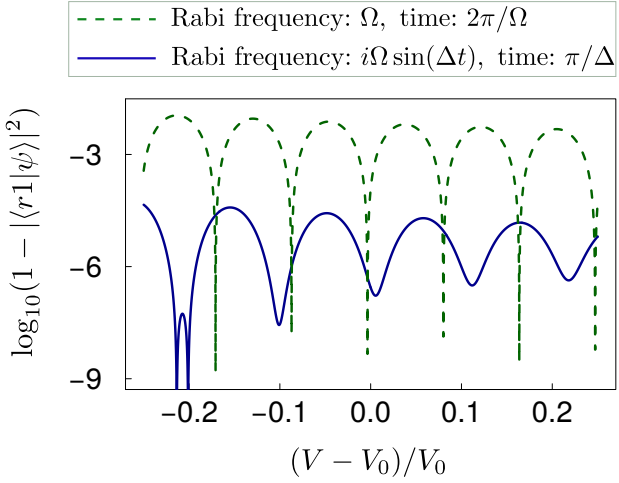}
\caption{Population loss in the the state $-i|r1\rangle$ after the second step of the Rydberg gate shown with the common logarithm and parameters $V_0:\Omega:\Delta=12:1:\pi^{-1}$. $V_0$ is the blockade interaction of $|rr\rangle$ when both qubits are located at the center of their traps; the actual Rydberg blockade $V$ can differ from $V_0$ due to the thermal motion of the qubits. The solid curve denotes results from propagating the wavefunction by the Hamiltonian in Eq.~(\ref{OurGate}) for a duration of $\pi/\Delta$. The dashed curve shows the population error of the state $-i|r1\rangle$ after an evolution time of $2\pi/\Omega$ under the Hamiltonian in Eq.~(\ref{TraditionalGate}). When averaged over all the $V$ in the figure, the population errors are $9.5\times10^{-6}$ and $3.7\times10^{-3}$ for the ORIR-based method and the traditional method, respectively. These two averages respectively become $1.7\times10^{-5}$ and $4.6\times10^{-3}$ if a larger interval of $(V-V_0)/V_0\in[-0.5,~0.5]$ is considered.   \label{figure05} }
\end{figure}

The population loss in the state $|r1\rangle$ after the second step of the gate is shown in Fig.~\ref{figure05} for different $(V-V_0)/V_0\in[-0.25,~0.25]$. The dashed curve in Fig.~\ref{figure05} shows that the blockade error in the traditional method is on the order of $10^{-3}$ for most values of $V$. Although at some special values of $(V-V_0)/V_0$~(for example, at around $-1/300$) a tiny population leakage of less than $10^{-5}$ can appear with the traditional method, the blockade error quickly rises to the level of $10^{-3}$ when $V$ deviates from that special value. Thus it is difficult to use the traditional method to achieve a small blockade error since it is challenging to sufficiently suppress the fluctuation of the qubit positions. On the other hand, the solid curve in Fig.~\ref{figure05} shows that the population loss in $|r1\rangle$ is much smaller when the second step is characterized by the effective Hamiltonian in Eq.~(\ref{OurGate}). In fact, the blockade error in the ORIR-based protocol of Fig.~\ref{figure05} is on the order of $10^{-5}$ for most values of $V$, which is two orders of magnitude smaller compared with the errors of the traditional method, effectively removing the blockade error.

The removal of blockade error by Eq.~(\ref{OurGate}) is quite different from other proposals of high-fidelity gates by time-dependent pumping. For example, comparing to the method of analytical derivative removal by adiabatic gate in Ref.~\cite{Theis2016}, our method only needs rectangular pulse (or quasi-rectangular pulse) that can be easily chopped from a continuous laser field with a pulse picker. Moreover, the ORIR-based gate does not have a strict dependence on specific blockade interaction for achieving the optimal fidelity, which can be found by comparing the solid curve in Fig.~\ref{figure05} for our method and Fig.~3 of Ref.~\cite{Theis2016} for the other method~[when comparing, we note that a fourth of $1-|\langle r1|\psi\rangle|^2$ in Fig.~\ref{figure05} contributes to the gate infidelity]. In fact, as an example, if the value of $V_0/\Omega$ in Fig.~\ref{figure05} increases to $30~(50)$, the population errors averaged over the interval $(V-V_0)/V_0\in[-0.25,~0.25]$ become $1.3\times10^{-6}$ and $5.9\times10^{-4}$~($4.3\times10^{-7}$ and $2.1\times10^{-4}$) in the ORIR-based method and the traditional method, respectively. These data show that the ORIR-based protocol can still suppress the blockade error by more than two orders of magnitude when $V/\Omega$ is large.

\begin{figure}
\includegraphics[width=3.2in]
{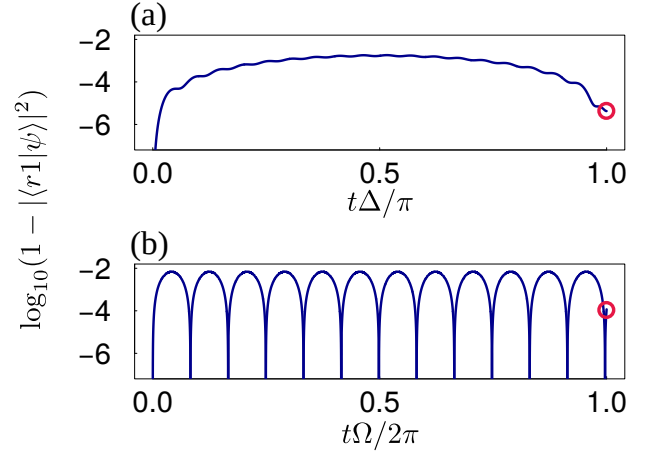}
\caption{Time evolution of $1-|\langle r1|\psi\rangle|^2$ during the second step of the Rydberg gate shown by the common logarithm with the condition of $V=V_0=12\Omega$. (a) shows the result from the ORIR-based method by using the Hamiltonian in Eq.~(\ref{OurGate}). (b) shows the results of the traditional method by using the Hamiltonian in Eq.~(\ref{TraditionalGate}). The red circles denote the population errors at the end of the second step of the gate, which are $4.3\times10^{-6}$ and $1.2\times10^{-4}$ in (a) and (b), respectively. \label{figure06} }
\end{figure}

To understand the mechanism for the suppression of the blockade error by ORIR, we show the time evolution of the population in $|rr\rangle$, which is equal to $1-|\langle r1|\psi\rangle|^2$, during the second step for our method and the traditional method in Fig.~\ref{figure06}(a) and~\ref{figure06}(b), respectively. Figure~\ref{figure06}(a) shows that the population leakage with ORIR increases with a minor oscillation pattern during the first half of the pulse, but decreases with a similar pattern during the latter half of the pulse. This removal of the blockade error is similar to the suppression of the blockade error with the spin-echo method studied in Ref.~\cite{Shi2018prapp2}, although the underlying physics is different.

\subsection{Gate error from finite edges and timing errors of laser pulses}\label{sec04A01}
As shown in Appendix~\ref{appendixA}, the finite rising and falling, and timing synchronization of laser pulses are also factors that limit the accuracy of the quantum gate. In most experiments on Rydberg quantum gates, an $s$- or $d$-orbital Rydberg state was used~\cite{Wilk2010,Isenhower2010,Zhang2010,Maller2015,Zeng2017,Picken2018,Levine2018,Graham2019,Levine2019}. Their excitation is usually via a two-photon excitation through a largely detuned intermediate state. In typical experiments, the laser beams for the upper transition are left on for several tens of nanoseconds longer than the lasers for the lower transitions~\cite{Maller2015}. Then, the onset and cutoff of the optical pumping are determined by the lasers of the lower transitions. So, we only need to consider four timing errors $\{t_{\text{ge}+}^{\text{(s)}},t_{\text{ge}-}^{\text{(s)}}, t_{\text{ge}+}^{\text{(e)}},t_{\text{ge}-}^{\text{(e)}}\}$ for the two lower laser beams, where $t_{\text{ge}\pm}^{\text{(s)}}$ and $t_{\text{ge}\pm}^{\text{(e)}}$ are the deviation of the actual start and cutoff times from the desired start and cutoff times of the pulse~(as illustrated in Fig.~\ref{timing-figure01}); ideally, they should be zero. These errors will influence both $|01\rangle$ and $|11\rangle$ among the four input states. According to the discussion in the fourth paragraph of this section and~\cite{Levine2018}, we choose $\Omega/2\pi=2$~MHz as the desired value of $\Omega_{\text{ge}\pm}$. As for the Rydberg blockade, we assume $V_0/\Omega=12$ and choose, as an example, $V = 0.97V_0$, where the population error in $|r1\rangle$ is about $10^{-5}$~($\approx$ average error for the solid curve in Fig.~\ref{figure05}) in the ideal case. The optical excitation occurs during $[0,~T]$ in the ideal case, where $T$ is the desired pulse duration. We suppose that the rising or falling of the laser pulses need $20$~ns~\cite{Urban2009,Maller2015,Levine2018}. The optimal value of $T$ corresponds to the smallest population error in $|01\rangle$. This is because the population leakage is less sensitive for the input state $|11\rangle$ since the transition $|r1\rangle\rightarrow|rr\rangle$ is blocked during the second step of the gate. With a numerically found optimal $T$ of $795.3963$~ns, the population leakage in $|01\rangle$~($|r1\rangle$) is $4.4~(7.7)\times10^{-6}$. We then use this $T$ to investigate the population errors in $|01\rangle$ and $|r1\rangle$ when the timing error is included. 

\begin{figure}
\includegraphics[width=3.2in]
{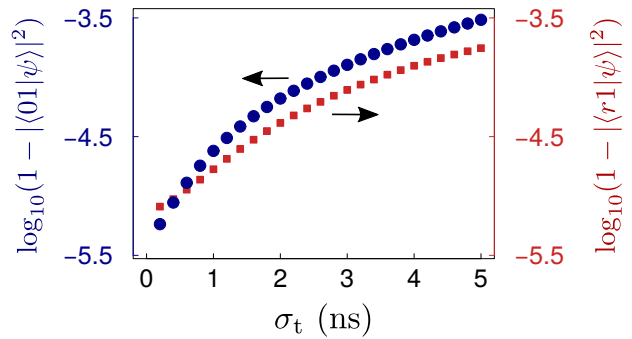}
\caption{Round~(square) symbols show the average population loss in the state $|01\rangle~(|r1\rangle)$ with a r.m.s. timing error $\sigma_{\text{t}}$ for the start and end of the two laser pulses. Calculation was based on propagating the state with the Hamiltonian in Eq.~(\ref{Hwithtimingerror}) during $[t_{\text{ge}\pm}^{\text{(s)}},~T+t_{\text{ge}\pm}^{\text{(e)}}]$. Average was calculated with Eq.~(\ref{6dimensionINT_app}) by letting $t_{\text{ge}\pm}^{\text{(s)}}$ and $t_{\text{ge}\pm}^{\text{(e)}}$ apply the values $\{-5\sigma_{\text{t}},-4\sigma_{\text{t}},\cdots,~5\sigma_{\text{t}}\}$. Here $(\Omega,~V,~\Delta)/2\pi=(2,~23.28,~2/\pi)$~MHz.        \label{figure07-new} }
\end{figure}

To include possible timing error in the second step with optical pumping for the target qubit, we cast Eq.~(\ref{OurGate}) into
\begin{eqnarray}
  \hat{H} &=&
  \left(\begin{array}{cc}
    V& \frac{\Omega_{\text{ge}+}}{4} e^{i\Delta t}-\frac{ \Omega_{\text{ge}-}}{4} e^{-i\Delta t} \\
 \frac{ \Omega_{\text{ge}+}}{4} e^{-i\Delta t}- \frac{\Omega_{\text{ge}-}}{4} e^{i\Delta t}  &0
    \end{array}
  \right) ,\nonumber\\
  \label{Hwithtimingerror}
\end{eqnarray}
where $\Omega_{\text{ge}\pm}$ is nonzero during $[t_{\text{ge}\pm}^{\text{(s)}},~T+t_{\text{ge}\pm}^{\text{(e)}}]$, and has a rising and falling edge during the first and last $20$~ns of the pulse. Each set of nonzero timing errors can result in gate error. Following Appendix~\ref{appendixA}, the average error is calculated with Eq.~(\ref{6dimensionINT_app}) where the summation runs through the $11^4$ sets of $\{t_{\text{ge}+}^{\text{(s)}},t_{\text{ge}-}^{\text{(s)}}, t_{\text{ge}+}^{\text{(e)}},t_{\text{ge}-}^{\text{(e)}}\}$. With the r.m.s. error $\sigma_{\text{t}}\in[0.2,~5]$~ns, the population losses in $|01\rangle$ and $|r1\rangle$ are shown by the round and square symbols in Fig.~\ref{figure07-new}, respectively. One can find that even with a very large fluctuation characterized by $\sigma_{\text{t}}=5$~ns, the lost population in $|01\rangle$~($|r1\rangle$) is only $3.0~(1.8)\times10^{-4}$ after the second step, which is much smaller than the blockade error in the traditional method even without pulse defects~(see the dashed curve in Fig.~\ref{figure05}). This suggests that ORIR is useful for high-fidelity quantum gate based on the mechanism of Rydberg blockade.


\section{Discussions}\label{sec05}

The ORIR-based time-dependent Rabi frequencies can also be used in trapped ions. Compared to neutral atoms, the entangling gates with trapped ions can attain a much better accuracy~\cite{Tan2015,Ballance2015,Ballance2016,Gaebler2016}. But compared to ions, neutral atoms are easy for building large-scale qubit arrays~\cite{Wang2015,Xia2015,Barredo2016,Wang2016}. Thus, it is useful to combine these two merits together in one system; indeed, there is intense interest to trap and manipulate Rydberg ions~\cite{Higgins2017,Higgins2017prl,Engel2018}, as well as theoretical effort to design protocols for quantum entanglement between Rydberg ions~\cite{Muller2008,Li2013,Li2014apb}. The methods in~\cite{Muller2008,Li2013,Li2014apb} depend on time-dependent Rabi frequencies and, in fact, the protocol in~\cite{Li2013} uses a Rabi frequency~$\propto\sin(\Delta t)$ that can be easily realized with quasi-rectangular laser fields chopped from continuous lasers with pulse pickers. Recently, adiabatic excitation of a single trapped Rydberg ion was experimentally demonstrated by using two sinusoidal Rabi frequencies in a ladder configuration~\cite{Higgins2017prl}. Reference~\cite{Higgins2017prl} reported that the reason to have used sinusoidal Rabi frequencies is that they result in a better fidelity than using Gaussian pulses. These results show that ORIR, which can offer sinusoidal Rabi frequencies by only using continuous laser and pulse pickers, is useful for Rydberg excitation in trapped ions.


\section{Conclusions}\label{sec06}
In conclusion, we have studied the application of an off-resonance-induced resonance~(ORIR) in single-site Rydberg addressing in a 3D lattice and high-fidelity Rydberg gate. These applications benefit from a time-dependent Rabi frequency $2\Omega \cos(\Delta t)$~[or $2i\Omega \sin(\Delta t)$] that naturally arises from the action of two symmetrically detuned coherent fields of equal~[or opposite] phase. ORIR can be implemented with quasi-rectangular laser pulse chopped from a continuous laser by an electro-optic modulator or an acousto-optic modulator, a method commonly used in experiments with Rydberg atoms.

In the study of quantum gates by Rydberg interactions of neutral atoms, ORIR can enable single-site Rydberg addressing in a 3D optical lattice while leaving irradiated nontarget atoms intact upon the completion of the Rydberg deexcitation of the target atom; moreover, we show that ORIR can suppress a fundamental rotation error in the Rydberg blockade gate, making it possible to achieve high fidelity with only quasi-rectangular pulses. Along the way, we find another method for single-site Rydberg addressing in a 3D lattice by using a ladder-type system. In this latter method, spin echo is used for any nontarget atom so that its state is restored when a single target atom is pumped to the Rydberg state, and the target atom picks up a $\pi$ phase when its state is restored to the ground state. These methods make it possible to entangle two selected qubits deep in a 3D atomic array and paves the way to large-scale quantum processors based on Rydberg atoms.


\section*{ACKNOWLEDGMENTS}
The author is grateful to Yan Lu and Tian Xia for fruitful discussions. This work was supported by the National Natural Science Foundation of China under Grant No. 11805146 and the 111 Project (No. B17035).

\appendix{}
\section{Transfer error from failure of timing synchronization of laser pulses}\label{appendixA}
The ORIR-based application in the quantum gate and single-site addressing depends on timing synchronization of laser pulses, which is discussed in this appendix.

\begin{figure}
\includegraphics[width=3.4in]
{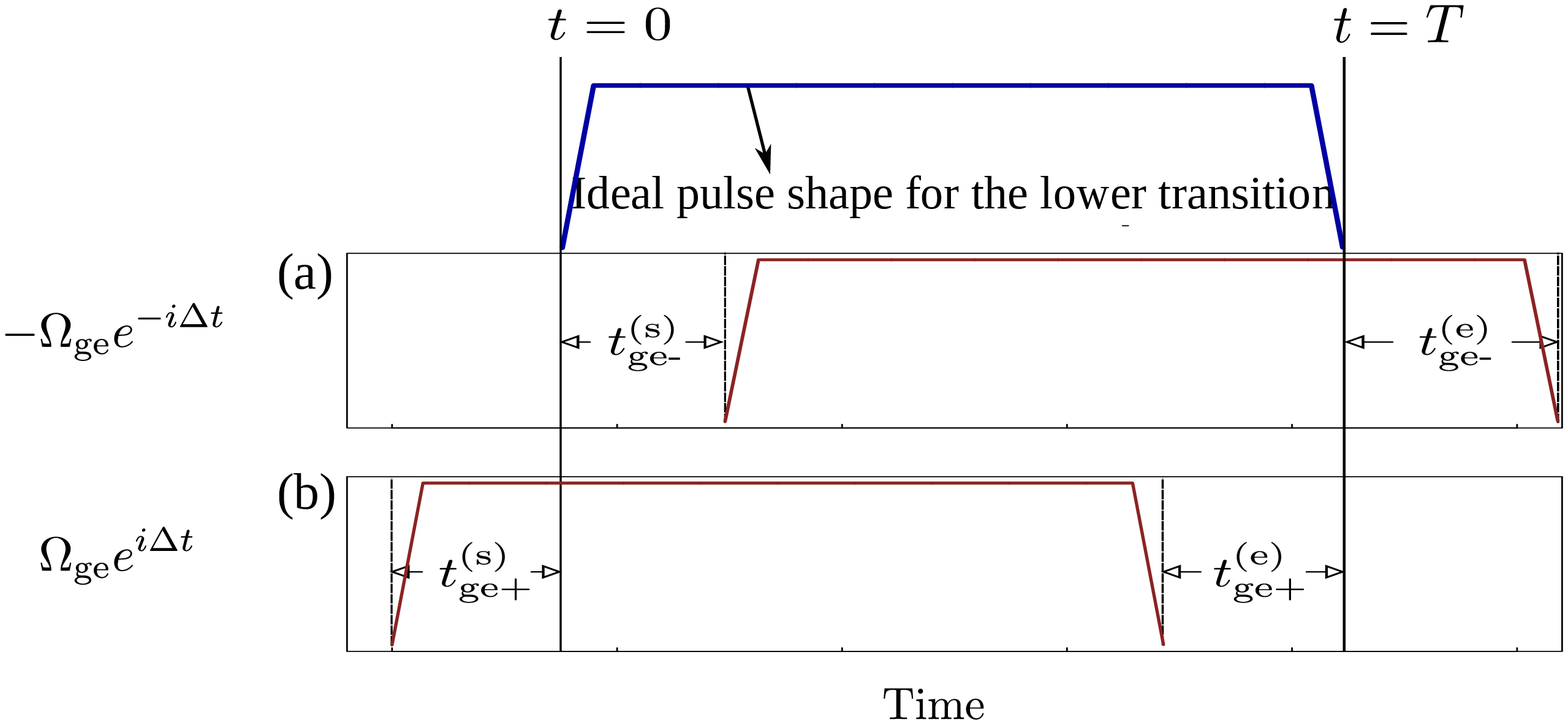}
\caption{Failure of timing synchronization of the laser pulses. The topmost curve shows the shape of the two lower laser pulses when they have no timing error. (a) and (b) show the pulse shapes with timing errors. \label{timing-figure01} }
\end{figure}

First, the timing synchronization of the laser pulses can be realized by using acousto-optic multi-channel modulators made with tellurium dioxide crystals, which can deflect multiple laser beams~\cite{mAOM}. To address arrays of rubidium atoms with two-photon excitation of Rydberg states, a Doppler-free multiwavelength~(780 and 480~nm) acousto-optic deflector was experimentally demonstrated about a dozen years ago~\cite{Kim2008}. The deflector in~\cite{Kim2008} can simultaneously diffract the two incident optical wavelengths with a common diffraction angle. So, it is possible to realize a multi-channel modulator for deflecting several laser lights. In this case, the durations of the two laser pulses in Fig.~\ref{timing-figure01} will be identical so that there will be no error due to failure of the timing synchronization.  

Second, the timing synchronization can be approximately achieved by using commercially available high-precision digital delay generators. In Ref.~\cite{Sabmannshausen2013}, a single delay generator (Quantum Composer 9528) was used to control all timings that were relevant for Rydberg-state excitation and detection. This type of delay generator has a resolution of 0.25~ns and r.m.s. jitter of 0.4~ns~\cite{QuantumComposer}, and is widely used in the experimental study on quantum physics~\cite{Matthews2011,Groote2015}, quantum chemistry~\cite{Zhu2019}, combustion~\cite{Eitel2017,Bonebrake2019}, accelerator~\cite{Smith2010}, and aerodynamics~\cite{Watkins2013}. The precision of commercial digital delay generators can also be surprisingly high. For example, the pulse generated by a laser synchronization module in~\cite{EKSPLA} has a resolution of 25~ps, an accuracy of 25 ps$+10^{-6}\times$delay, and jitter of less than 30~ps. However, ultra-high precision is not required for most applications, so that such digital delay generators are not well-known. To use these digital delay generators for timing, the transmission time of the control signal should also be considered, which is possible by adjusting the length of the cable.  

We suppose the pulse timing is controlled by a commercial digital delay generator and assume that the r.m.s timing errors of the onset and cutoff of the pulse are both $\sigma_{\text{t}}$. The beginning of the pulse is at a time $t$ that obeys Gaussian distribution $f(t|0,\sigma_{\text{t}}) = e^{-t^2/(2\sigma_{\text{t}}^2)}/\sqrt{2\pi\sigma_{\text{t}}^2}$, while the end of the pulse is at a time $t$ that is distributed according to $f(t|T ,\sigma_{\text{t}}) = e^{-(t-T )^2/(2\sigma_{\text{t}}^2)}/\sqrt{2\pi\sigma_{\text{t}}^2}$, where $T$ is the desired pulse duration. The timing diagram of the four laser beams is shown in Fig.~\ref{timing-figure01}. Altogether, there are four timing errors, i.e., the two timing errors of the pulse arrival, $t_{\text{ge}\pm}^{\text{(s)}}$, of the two laser lights for the lower transition, and their cutoff, $t_{\text{ge}\pm}^{\text{(e)}}$. The fidelity $\mathcal{F}$ for each realization of the gate is a function of these four timing errors. Then, the expected fidelity is given by
\begin{eqnarray}
  \overline{\mathcal{F}}   &=& \int \mathcal{F}(t_{\text{ge}+}^{\text{(s)}},t_{\text{ge}-}^{\text{(s)}}, t_{\text{ge}+}^{\text{(e)}},t_{\text{ge}-}^{\text{(e)}})\nonumber\\
  &&\cdot \mathcal{P}(t_{\text{ge}+}^{\text{(s)}},t_{\text{ge}-}^{\text{(s)}}, t_{\text{ge}+}^{\text{(e)}},t_{\text{ge}-}^{\text{(e)}})  \mathscr{D}_4,\label{6dimensionINT}
\end{eqnarray}
where 
\begin{eqnarray}
 \mathcal{P}  &=& f(t_{\text{ge}+}^{\text{(s)}}|0,\sigma_{\text{t}})f(t_{\text{ge}-}^{\text{(s)}}|0,\sigma_{\text{t}})f(t_{\text{ge}+}^{\text{(e)}} |T ,\sigma_{\text{t}}) f(t_{\text{ge}-}^{\text{(e)}} |T ,\sigma_{\text{t}}),\nonumber
\end{eqnarray}
and $\mathscr{D}_4$ indicates a four-dimensional integral given by $dt_{\text{ge}+}^{\text{(s)}}dt_{\text{ge}-}^{\text{(s)}}d t_{\text{ge}+}^{\text{(e)}}dt_{\text{ge}-}^{\text{(e)}}$. Each value of $\mathcal{F}$ in Eq.~(\ref{6dimensionINT}) is evaluated by integration of the Schr\"{o}dinger equation through the pumping process; this results in quite a long time to evaluate the four-dimensional integration in Eq.~(\ref{6dimensionINT}). Since the timing errors obey a normal distribution, the contribution from large timing errors is negligible. For this reason, we approximate Eq.~(\ref{6dimensionINT}) by considering eleven values $\in\{\pm5\sigma_{\text{t}},\pm4\sigma_{\text{t}},\pm3\sigma_{\text{t}},\pm2\sigma_{\text{t}},\pm\sigma_{\text{t}},0\}$ for the integration over each of the four timing errors, so that 
\begin{eqnarray}
 \overline{\mathcal{F}}   &\approx&\frac{\sum \mathcal{F} \mathcal{P} }{\sum  \mathcal{P}},\label{6dimensionINT_app}
\end{eqnarray}
where `$\sum$' sums over the $11^4$ sets of $\{t_{\text{ge}+}^{\text{(s)}},t_{\text{ge}-}^{\text{(s)}}, t_{\text{ge}+}^{\text{(e)}},t_{\text{ge}-}^{\text{(e)}}\}$.


%

\end{document}